\RequirePackage{fix-cm}
\documentclass[smallextended]{svjour3}       % onecolumn (second format)
\smartqed  % flush right qed marks, e.g. at end of proof
\usepackage{graphicx}
\usepackage{color}
\usepackage{soul}
\usepackage{amsmath}
\usepackage{amssymb}
\usepackage{url}

\journalname{J Stat Phys}
\begin{document}

\title{Value production in a collaborative environment \thanks{Supported by EU's FP7 FET Open STREP Project: ICTeCollective No. 238597.
%Grants or other notes
}
}
\subtitle{Sociophysical studies of Wikipedia}


\author{Taha~Yasseri         \and
	J\'{a}nos~Kert\'{e}sz %etc.
}
%\authorrunning{} % if too long for running head

\institute{T.~Yasseri  \at
              Department of Theoretical Physics,
	      Budapest University of Technology and Economics, Budapest, Hungary.\\
	     \email{yasseri@phy.bme.hu}             \\
             \emph{Present address: Oxford Internet Institute, University of Oxford, Oxford, United Kingdom}
              \and	    
	   J.~Kert\'{e}sz \at
 Center for Network Science, Central European University, Budapest, Hungary,\\
 and Department of Theoretical Physics,
	      Budapest University of Technology and Economics, Budapest, Hungary.\\
	      \email{janos.kertesz@gmail.com}           %  \\           
}

\date{Received: date / Accepted: date}

%%%%%%%%%%%%%%%%%%%%%%%%%%%%%%%%%%%%%%%%%%%%%%%%%%%%%%%%%%%%%%%%%%%%%%%%%%%%%%%%%%%%%%%%%%%%%%%%%%%%
\maketitle

\begin{abstract}
We review some recent endeavors and add some new results to characterize and understand underlying mechanisms in Wikipedia (WP), 
the paradigmatic example of collaborative value production. We analyzed the statistics of editorial activity 
in different languages and observed typical circadian and weekly patterns, which enabled us to estimate the 
geographical origins of contributions to WPs in languages spoken in several time zones. Using a recently 
introduced measure we showed that the editorial activities have intrinsic dependencies in the burstiness of events. 
A comparison of the English and Simple English WPs revealed important aspects of language complexity and showed how peer cooperation solved the task of enhancing readability. 
One of our focus issues was characterizing the conflicts or edit wars in WPs, which helped us to  
automatically filter out controversial pages. When studying the temporal evolution of the controversiality of 
such pages we identified typical patterns and classified conflicts accordingly. Our quantitative analysis provides 
the basis of modeling conflicts and their resolution in collaborative environments and contribute to the understanding 
of this issue, which becomes increasingly important with the development of information communication technology.


%\PACS{PACS code1 \and PACS code2 \and more}
%\subclass{MSC code1 \and MSC code2 \and more}
\end{abstract}
%%%%%%%%%%%%%%%%%%%%%%%%%%%%%%%%%%%%%%%%%%%%%%%%%%%%%%%%%%%INTROINTROINTROINTORINTROINTROINTROINTOR
\section{Introduction}\label{sec:intro}

Wikipedia (WP) is a truly amazing product of the 21st century. It is a free online 
encyclopedia \footnote{\url{http://en.wikipedia.org/wiki/Wikipedia}}
edited by volunteers, which has achieved within short period of time enormous success: This encyclopedia, 
which practically anyone can contribute to has a comparable reliability to the highly professional Encyclopedia 
Britannica \cite{giles2005} and has got by now the number one general work of reference in everyday practice. 
The main question related to Wikipedia is: How can an encyclopedia be reliable if anyone can edit it? The 
{\it bon mot} of Wikipedians is not a satisfactory answer, namely that ``It works only in practice.  In theory, 
it can never work.''

The literature about WP is overwhelming. Without seeking completeness, Okoli et~al. \cite{okoli2012} tracked 
more than 2000 related articles. However, there are rather comprehensive reviews,
e.g., \cite{nielsen2011,jullien2012} and an overview of the 
visibility of WP in scholarly publications \cite{park2011}.  
In addition, there are also online platforms to collect and index WP-related 
academic literature; among them are ``WikiLit''\footnote{\url{http://wikilit.referata.com/wiki/Main_Page}} \cite{ayers2011},
 ``AcaWiki''\footnote{\url{http://acawiki.org/Home}}
and ``WikiPapers'' \footnote{\url{http://wikipapers.referata.com/wiki/Main_Page}}.
A monthly review of the most recent scholarly studies on WP is also available 
at ``Wikimedia research Newsletter''.\footnote{\url{http://meta.wikimedia.org/wiki/Research:Newsletter}}

First Wikipedia studies were mostly on its size and growth, showing an 
initial exponential growth \cite{voss2005,almeida2007}, 
which was later reported to be saturating by other authors \cite{suh2009}.
Another main line of WP research is focused on vandalism 
detection \cite{smets2008,potthast2008,wu2010,west2010,adler2011}. 
Assessing user reputation \cite{adler2006,Javanmardi2010b} and investigating the articles
quality \cite{hu2007,wilkinson2007,luyt2008,stvilia2008,jones2008,kittur2008,javanmardi2010a,wu2011} are 
other two important topics.
To understand the management system of WP, there have been interesting studies 
on user authority, adminship, 
governance and promotion strategy \cite{leskovec2010,aaltonen2011,derthick2011,sepehrirad2012,lee2012},
 in addition to analysis of WP 
policies and bureaucracy \cite{butler2008}. A considerable amount of WP policies 
are on what to be/not to be in WP. Consequently, there are studies on
topical coverage and notability of entries \cite{holloway2007,halavais2008,taraborelli2010}.
Seeing WP as a network of articles, various researchers offer 
analysis and models for topology and growth of the
Wikigraph \cite{ratkiewicz2010a,capocci2006,buriol2006,zlatic2006,zlatic2011}, 
whereas some others used WP to build up
knowledge taxonomies and semantic structures
\cite{strube2006,muchnik2007,ponzetto2007,capocci2008,zesch2008,kittur2009,silva2011,suchecki2012}.
Masucci et al. showed that semantic space has a scale-free structure by analyzing 
information extracted from WP \cite{masucci2011}.
More to the sociological side, Restivo and van de Rijt studied the effect of 
social awards on users activity \cite{restivo2012} and 
Lam et~al. explored the gender imbalance among WP editors \cite{lam2011}.
Massa presented an algorithm to extract the social network of editors \cite{massa2011}, and  
Danescu-Niculescu-Mizil et~al. have studied talk page conversations to 
observe the relation between language coordination
and social power of editors \cite{mizil2012}.
Clearly, scholarly studies in the field of peer-production go beyond WP and for instance 
Roth et~al. studied dynamics of communities of wiki-based projects in
the whole ``WikiSphere'' \cite{roth2008}. Finally, in a rather different approach, Mesty\'an et al. have made use of WP
edit and page view data to predict the movies box office revenue \cite{mestyan2012}.

Our motivation to study Wikipedia comes from the need of understanding the laws of modern collaborative 
value production. This is of great importance as in our increasingly complex world the role of information 
communication technology (ICT) mediated peer collaboration is expected to become more and more important in 
the future. Due to its relation to  ICT, the methods of "Computational Social Science" (CSS) \cite{lazer2009} 
are adequate tools of investigation of such collaborations. CSS is a truly multidisciplinary endeavor with a 
considerable contributions from physicists (see, e.g., \cite {chakrabarti2006}). The main difference between traditional 
social science and CSS is that the latter is data driven: it uses the digital footprints we leave behind in almost 
all our activities in the digital era \cite{bohannon2006}. 

Collaboration has always been fundamental to most human achievements. Modern information communication technology 
opens up entirely new ways of cooperation, where partners can interact remotely with an unprecedented speed exchanging 
extremely large amount of information. Tim Barners-Lee developed originally the World Wide Web \cite{wiki:www} at CERN in 
order to create an appropriate platform for huge collaborations, which are  ubiquitous in high energy physics. Another 
important example is that of free software development as defined by the Free Software Foundation\footnote{\url{http://www.fsf.org/}}. Nowadays 
all major scientific projects from the Human Genome\footnote{\url{http://www.ornl.gov/sci/techresources/Human_Genome/project/about.shtml}} to Hubble Space Telescope\footnote{\url{http://www.nasa.gov/mission_pages/hubble/story/the_story_2.html}} rely heavily on ICT 
mediated collaboration but even on smaller scale we often use Current Version Management\footnote{\url{http://savannah.nongnu.org/projects/cvs}}, wiki \cite{leuf2001} and 
related environments to increase efficiency. WP is a paradigmatic example of collaborative environment with the additional 
advantage that all the changes and interactions are well documented and publicly available, which makes it particularly suitable 
for scientific studies. 

Many questions arise when studying WP from our point of view. What are the characteristic features of editorial activities? 
How are they related to other examples of human dynamics, which have been intensively studied in CSS \cite {barabasi2005,karsai2012}? 
What is the mechanism behind the emergence of an article? How can the complexity of the product of the cooperation, namely that of the 
articles be characterized? How do conflicts emerge and get resolved? In the following we will present analysis of WP data in order to 
contribute to the clarification of these questions. 

To accustom the unfamiliar readers to the terminology and work-flow of Wikipedia, 
in the next Section we briefly review main tools and objects in the Wikimedia platform. 
Familiar readers are encouraged to skip this Section. In Section~\ref{sec:methods}, 
we explain different methods and sources for collecting WP data, and in Section~\ref{sec:results},
a summary of our recent \cite{sumi2011a,sumi2011b,yasseri2012a,yasseri2012b,yasseri2012c,torok2012} and some new results
is provided and compared to the related reports by other authors.
We close the paper with a conclusion (Section~\ref{sec:conclusion}).

\section{How Wikipedia works}
%Before going to more details about the methods and results of our studies on WP, we give a short summary of the editorial process in WP and describe the basis, on which WP works on. 

WP has more than 280 language editions at the moment. Main concepts 
and structures are similar in all language editions with little variations due to local
modifications by the editors' community of the specific edition. Later we will deal with several WPs, however, whenever it is not specified else explicitly, the English WP is meant.

Describing the structure of WP, there are two main elements to name,
i) Articles ii) WP editors, also called ``Wikipedians''.
The rest is all about the internal and inter-element connections and interactions
of the members of these two groups, which we name ``Accessories'' in this
paper.

\subsection{Articles}\label{sec:article}
Wikipedia, similarly to any other encyclopedia consists of entries about 
different topics, hereafter called Articles. Each article of WP
has necessarily a ``title'', a nonempty ``content'', and a ``history'' which 
is a collection of all previous revisions of the article beginning 
from its inception. 
In Table~\ref{tab:article}, some basic statistics of articles in the ten largest WPs are given.
\begin{table}
\caption{Article statistics for 10 largest Wikipedias. First and second columns are indicating the Language and the Symbol of the Wikipedia editions. 
In the following columns number of Articles (divided by 1000), Average Length of the articles in characters, number of editors with at least one edit, divided by the number of articles,
 and number of Featured articles are reported.}
\label{tab:article}    
\begin{tabular}{llllllll}
\hline\noalign{\smallskip}
Language	&	Symbol	&	Art. (k)	&	Av. Len.	&	Av. Edit/Art.	&	Editor/Art.	&	Featured	\\
\noalign{\smallskip}\hline\noalign{\smallskip}													
English	&	en	&	4,080	&	5544	&	136.7	&	1.26	&	3638	\\
German	&	de	&	1,454	&	5081	&	77.3	&	0.39	&	2113	\\
French	&	fr	&	1,287	&	5189	&	68.0	&	0.32	&	1093	\\
Dutch	&	nl	&	1,072	&	2567	&	30.6	&	0.14	&	272	\\
Italian	&	it	&	957	&	4799	&	59.5	&	0.20	&	538	\\
Polish	&	po	&	917	&	3634	&	35.6	&	0.15	&	530	\\
Spanish	&	es	&	915	&	5027	&	69.2	&	0.53	&	1035	\\
Russian	&	ru	&	892	&	7913	&	59.1	&	0.24	&	553	\\
Japanese	&	ja	&826	&	6357	&	54.6	&	0.32	&	66	\\
Portuguese	&	pt	&739	&	3421	&	43.5	&	0.26	&	694	\\
\noalign{\smallskip}\hline
\end{tabular}
\end{table}
Articles of each language edition are connected via internal links. That makes the whole 
language edition a directed  
graph. Ideally this graph should be 
connected, 
however there are always ``Orphan'' (not linked by any article) and ``Dead-end'' (not linking 
to any other article) articles. There are also inter-language links, connecting
articles from different language editions. 

In general, articles could be edited by any Internet user. However there are protections 
against vandalism applied to some articles and prohibiting different classes of editors
from editing. Access to more complex actions, e.g. creating a new article, 
changing the title, or deleting an article is also subject to hierarchal structure of
editors (described in the next Section).

\paragraph{Featured articles:} ``Featured articles are considered to be the best articles 
Wikipedia has to offer, as determined by Wikipedia's 
editors.''\footnote{\url{http://en.wikipedia.org/wiki/Wikipedia:Featured_articles}}
Articles are tagged as featured based on the community decision on their 
accuracy, neutrality, completeness and style. In English WP there are more than
3,500 featured articles (see Table~\ref{tab:article}).

\paragraph{Lists of controversial articles:} There are also lists of articles with 
severe editorial disagreements in their history, see, e.g., ``List of Controversial Articles''
\footnote{\url{http://en.wikipedia.org/wiki/List_of_controversial_articles}}, and 
List of ``Lamest Edit Wars''\footnote{\url{http://en.wikipedia.org/wiki/Wikipedia:Lamest_edit_wars}}.
However, the accuracy and coverage of those lists are questionable. There is no clear 
definition and systematic algorithm to determine, which articles should be listed.

\subsection{Wikipedians}\label{sec:editors}

In principle any person with access to Internet could be a Wikipedia editor. 
Editors are recognized by the system based on the
IP addresses,  through which they are connected or with their user-name which they choose upon registration. As long as editors edit via their user-names, in general
no personal information about them is revealed, unless voluntary disclosure by 
themselves. There are semi-annual surveys run by Wikimedia Foundation to provide some
demographical information about the community of WP 
editors.\footnote{\url{http://meta.wikimedia.org/wiki/Editor_Survey_2011}} However, 
since participation in the survey is completely voluntary, the 
reliability and coverage of this information is questionable. Therefore, personal information of the editors'
community of WP, is the most unknown aspect of it.

There is a well defined hierarchal structure among Wikipedians, 
such that editors from different classes have access to certain editorial actions. For exact description of each level rights and accessed, 
see \url{http://en.wikipedia.org/wiki/Wikipedia:User_access_levels}. In brief, some of these classes, common in all language editions are:
a) Unregistered users, with the right to edit unprotected existing pages. 
b) New users, with the right to edit unprotected pages and create new pages.  
c) Auto-confirmed users, with the right to edit semi-protected pages and move pages to new titles.
d) Administrators (admins), with the right to edit protected pages, delete or protect pages, and block other editors from editing,
c) Bureaucrats, with the right to change the user rights and in most of the Wikipedias, conclude the promotion polls.  
Promotion to higher levels, starting from adminship, is upon decision of the editors' community confirmed by promotion polls.
In Table.~\ref{tab:user}, some basic statistics on the editors' communities of 10 largest language editions are given.
\begin{table}
\caption{Editor statistics for 10 largest Wikipedias. First and second columns are indicating the Language and the Symbol of the Wikipedia editions. 
In the following columns, number of Registered users, users who have actually Contributed (at least one edit), Administrators, Bureaucrats, and the editors who are banned forever,
are reported.}
\label{tab:user}    
\begin{tabular}{lllllll}
\hline\noalign{\smallskip}
Language& Symbol& Registered & Contributed& Admins & Bureaucrats & Banned  \\
\noalign{\smallskip}\hline\noalign{\smallskip}
English & en  & 17,186,079 &5,085,719& 1,461 & 34 &93,812\\
German & de  & 1,467,633 & 564,993&268 &5  &7,978\\
French & fr  & 1,332,309 & 413,091&195 &7  &3,214\\
Dutch & nl  & 469,358 & 147,758&64 &9  &1,011\\
Italian & it  & 773,446 & 192,198& 105  & 6 &2,995\\
Polish & po  & 501,381 & 138,804&157  & 6 &731\\
Spanish & es  & 2,292,694 & 485,802 &134  & 133 &5,473\\
Russian & ru  & 886,133 &215,759 & 92  & 5 &2,855\\
Japanese & ja  & 643,770 &260,206 & 61  & 9 &5,693\\
Portuguese  & pt  & 1,026,749 & 198,000&37  & 7 &1,150\\
\noalign{\smallskip}\hline
\end{tabular}
\end{table}

\subsection{Accessories}\label{sec:access}
Around half of the edits in Wikipedia are outside the main name-space, i.e. in accessory pages \cite{kittur2007}.
This pages control the underlying mechanism of growth and maintenance of WP articles in the main name-space.
Here we briefly describe some of them.
\paragraph{Policies, guidelines, essays and instructions:}
``Wikipedia's policies and guidelines are pages that serve to document 
the good practices that are accepted in the Wikipedia 
community.''\footnote{\url{http://en.wikipedia.org/wiki/Wikipedia:Policies_and_guidelines}}
These policies are however subject of change and improvement by the community of editors and 
may slightly differ among different language editions.
\paragraph{User pages:}
``User pages are for communication and collaboration.''\footnote{\url{http://en.wikipedia.org/wiki/Wikipedia:User_pages}} 
They could be used to provide personal information of the editor or less encyclopedic content related to the editor.
 However, as they are part of the encyclopedia project, their content should not violate the main guidelines.
\paragraph{Article talk pages:}
The purpose of a Wikipedia talk page  is to provide space for 
``editors to discuss changes to its associated article or project 
page''.\footnote{\url{http://en.wikipedia.org/wiki/Wikipedia:Talk_pages}}
Talk pages are the main channels for social interactions between editors, and 
supposed to be the main place to resolve disagreements and editorial conflicts.
\paragraph{User talk pages:}
User talk pages are designed for more general communications directly to each 
editors. User talk pages are usually less technical than article talk pages
and conversations are more personal.
\paragraph{Common discussion pages:}
apart from article and user talk pages, there other discussion pages related to 
specific projects, polls, and more collective activities.
There are also different communication channels for Wikipedians outside of the 
WP, e.g., IRC channels and Wikimedia mailing lists; for an overview see \cite{pentzold2006}.
\paragraph{Categories:}
Categories are intended to group together pages on similar 
subjects.\footnote{\url{http://en.wikipedia.org/wiki/Wikipedia:Categorization}}
Categories are a feature of the MediaWiki platform. The latter allows articles to 
be grouped and provides the facility for the readers to 
navigate through the related  articles.
The process of article categorization, is carried out by editors, and 
its accuracy is at the same level as other content of WP.

\begin{table}
\caption{Page statistics for 10 largest Wikipedias. First and second columns are indicating the Language and the Symbol of the Wikipedia editions. 
In the following columns, number of All pages, Articles, Article Talk pages, User Pages, User Talk pages, and Categories, and  in the last
column sum of number of  Wikipedia guidelines, projects, polls and Help pages are reported. All the numbers are divided by 1000.}
\label{tab:page}    
\begin{tabular}{lllllllll}
\hline\noalign{\smallskip}
Language	&	Sym.	&	All	&	Art.	&	Art. Talk	&	Us. Page	&	Us. Talk	&	Cat.	&	WP, Help	\\
\noalign{\smallskip}\hline\noalign{\smallskip}																	
English	&	en	&	28,068	&	4,080	&	4,212	&	1,527	&	8,057	&	891	&	737	\\
German	&	de	&	4,062	&	1,454	&	458	&	338	&	342	&	153	&	35	\\
French	&	fr	&	5,268	&	1,287	&	1,024	&	166	&	951	&	213	&	33	\\
Dutch	&	nl	&	2,320	&	1,072	&	73	&	107	&	456	&	70	&	15	\\
Italian	&	it	&	3,045	&	957	&	188	&	64	&	851	&	172	&	96	\\
Polish	&	pl	&	1,766	&	917	&	219	&	73	&	95	&	101	&	26	\\
Spanish	&	es	&	3,845	&	915	&	184	&	125	&	971	&	184	&	23	\\
Russian	&	ru	&	3,053	&	892	&	362	&	77	&	245	&	212	&	28	\\
Japanese	&	ja	&	2,232	&	826	&	163	&	75	&	332	&	100	&	77	\\
Portuguese	&	pt	&	2,994	&	739	&	378	&	70	&	922	&	145	&	58	\\
\noalign{\smallskip}\hline
\end{tabular}
\end{table}

%%%%%%%%%%%%%%%%%%%%%%%%%%%%%%%%%%%%%%%%%%%%%%%%%%%%%%%%%%METHODSMETHODSMETHODSMETHODSMETHODSMETHODS
\section{Methods and Data}\label{sec:methods}
Beyond usual statistical methods to study Wikipedia, there are numerous open source software 
packages for different analyzing tasks.
Among them is ``WikiTrust''\footnote{\url{http://www.wikitrust.net/}} \cite{adler2006}, to 
measure article quality and assign a reputation to it. 
WikiXray\footnote{\url{http://meta.wikimedia.org/wiki/WikiXRay}} \cite{ortega2009} is another 
package for doing in-depth statistical analysis on different parameters, e.g. size of WPs, 
size of articles, number of contributers to each article, 
etc. However, since all WP data is publicly available, developing home made packages to 
analyze this data is a common approach.

\subsection{Data}
Every single action of Wikipedia editors is tracked and recorded. 
This includes all edits on articles, posts on talk pages, page deletions or creations,
changes in page titles, uploading multimedia files, etc.
Apart from the practical advantages of this complete archiving, it is also extremely 
valuable from scientific point of view. WP is one of the few
human societies that the history of all actions of its members are recorded and 
accessible.\footnote{Except deleted revisions, which are only available for admins and higher, 
and ``overseen'' revisions, which are accessible by no one.} 

\paragraph{Live data:} 
There are two convenient ways to access live data of Wikipedia. i) ``Wikimedia 
Toolserver\footnote{\url{http://toolserver.org}} 
databases, which contains a replica of all Wikimedia wiki databases, and ii) ``MediaWiki web 
service API''\footnote{\url{https://www.mediawiki.org/wiki/API}}.
For statistical analysis of contributions, Toolserver database tables are among the best 
sources of information.

\paragraph{Dumped data} 
Wikipedia also offers archived copies of its content in different formats\footnote{\url{http://dumps.wikimedia.org}}, 
e.g., XML and HTML and different types, e.g., 
snapshots of full history of articles or a collection of latest version of all articles. 
Generally for historical text analysis  of  articles,
the most reliable source would be these static copies.

\paragraph{Semantic Wikipedia} 
``Semantic Wikipedia'', as a general concept would be a combination of Semantic Web and WP 
data to provide structured data sets through query services.
There are various projects providing access to Semantic WP.
Examples are ``DBpedia''\footnote{\url{http://dbpedia.org}} \cite{auer2007},
``Semantic MediaWiki''\footnote{\url{http://semantic-mediawiki.org}} \cite{volkel2006}, and
``Wikipedia XML corpus''\footnote{\url{http://www-connex.lip6.fr/~denoyer/wikipediaXML}} \cite{denoyer2006}.
For a list of Semantic WP projects see \url{http://en.wikipedia.org/wiki/Wikipedia:Semantic_Wikipedia}.

%%%%%%%%%%%%%%%%%%%%%%%%%%%%%%%%%%%%%%%%%%%%%%%%%%%%%%%%%RESULTSRESULTSRESULTSRESULTSRESULTSRESULTS 
\section{Results and Discussion}\label{sec:results}

\subsection{Editorial habits}\label{sec:habit}
Similarly to any other large human society, the community of Wikipedia editors is very inhomogeneous. 
Editors vary in age, gender, nationality, education, occupation,
religion, interest, etc. 

\subsubsection{Edits statistics}\label{sec:stat}
\begin{figure} \sidecaption
\includegraphics[width=0.75\textwidth]{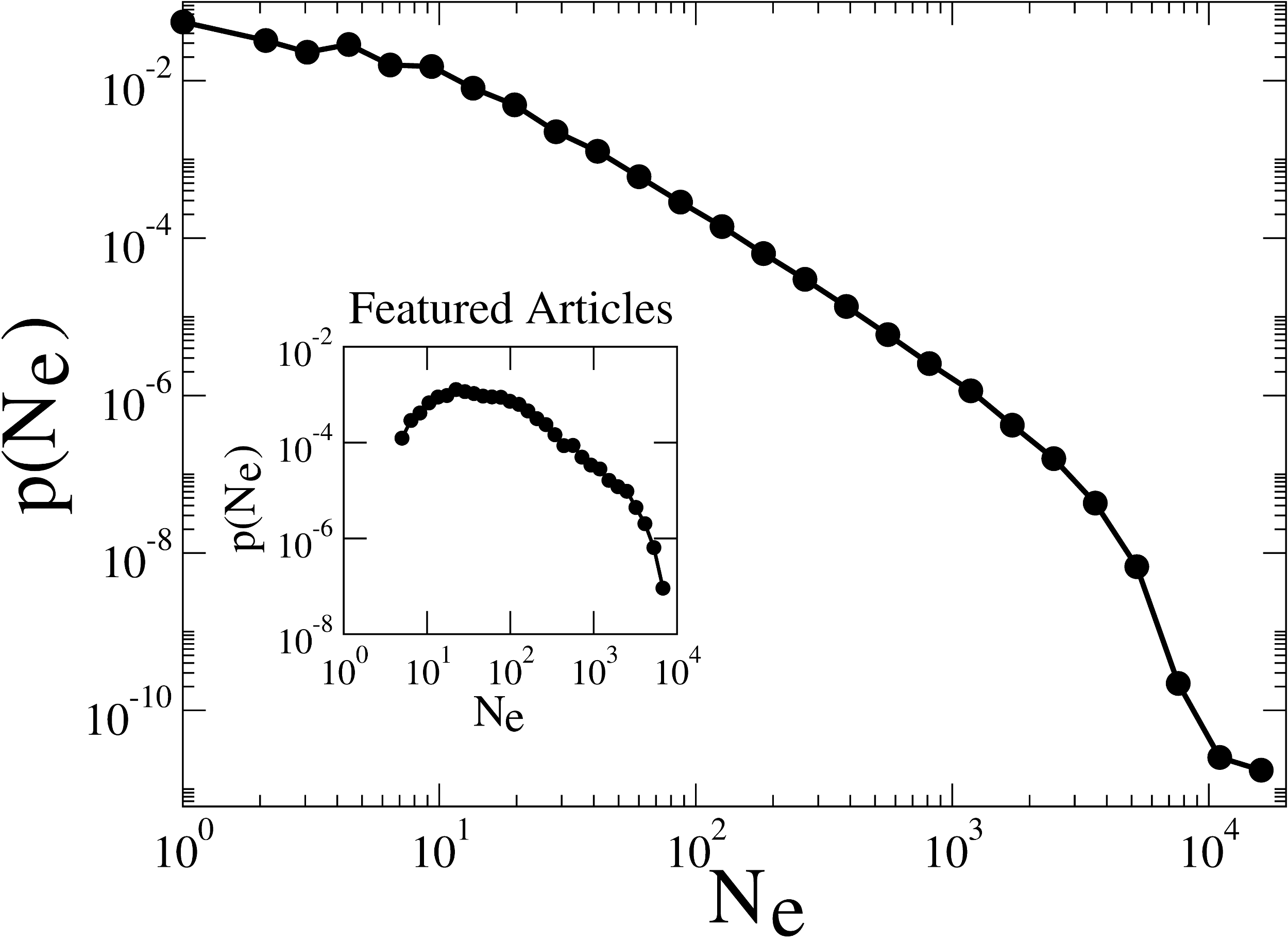} 
\caption{Probability distribution function of number of different human-editors 
contributed to each article. Most of the articles are basically edited by few editors, 
and few of them are edited by a large editorial pool of 10,000 editors. 
{\it Inset:} Probability distribution function of number of editors of the 3,122 ``featured articles''. The lower cut off of the distribution
moves from its natural value of 1 for the whole sample, to a value of 5 editors at least in 
the featured articles sample. The peak of the distribution
is about 22 editors, which could be considered as the optimal number of editors to achieve an acceptable quality for an
article.} \label{fig:user-article}
\end{figure}

Heterogeneity is present in the level of activity. Not only the total number of edits by 
each editor has a largely extended distribution \cite{wilkinson2008,ortega2008,javanmardi2009}, 
but also the number of different articles each editor contributes to is varying considerably from one 
editor to another \cite{ortega2009}. Finally, the number of editors contributing to an article  
has also a fat-tailed distribution (see Fig.~\ref{fig:user-article}), however, with a lower cut-off when we only consider
a selection of ``Featured Articles'', 
which are supposed to be articles with high level of completeness and accuracy.

\begin{figure} \sidecaption
\includegraphics[width=0.5\textwidth]{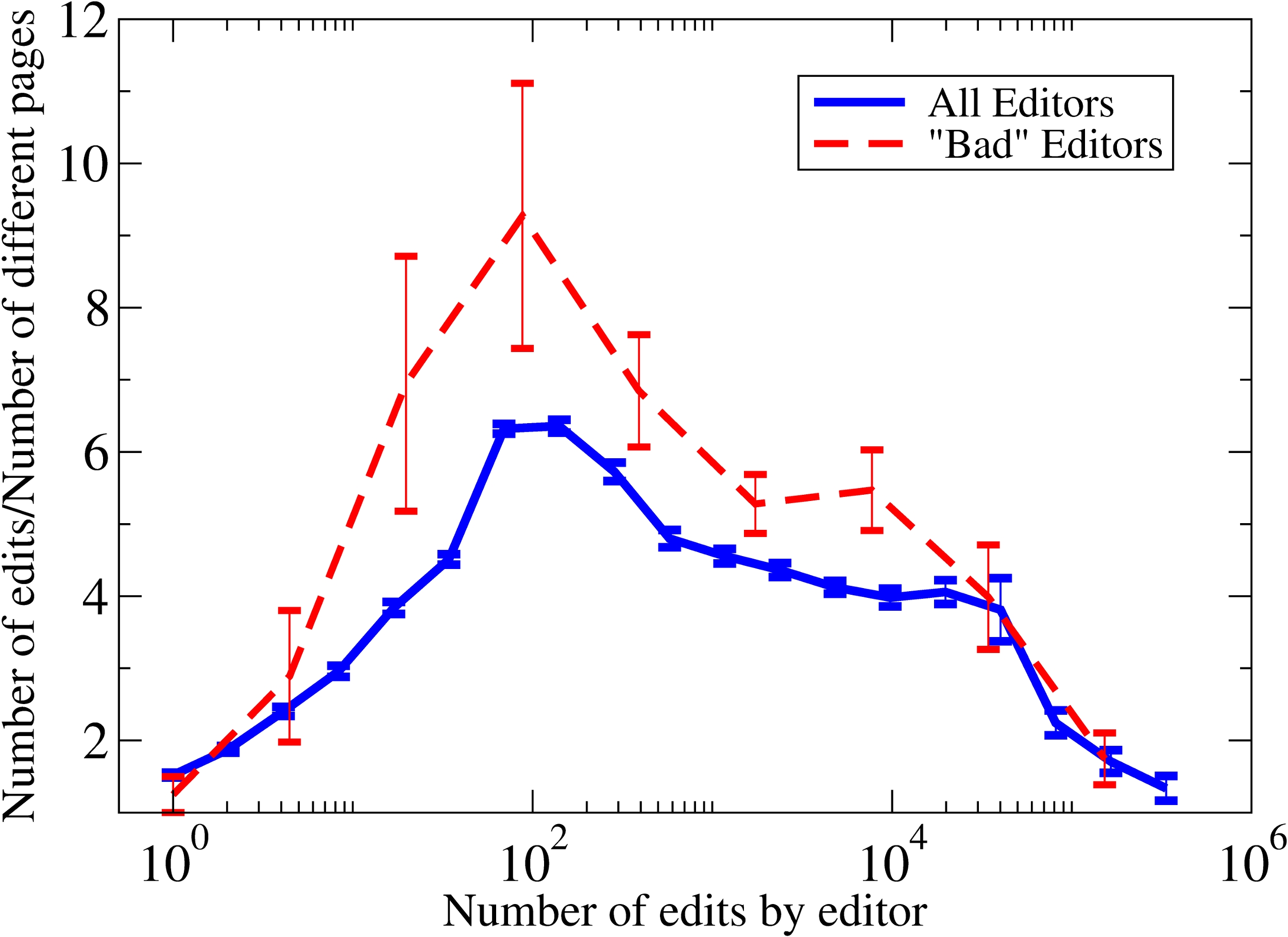} 
\caption{Average number of edits per article vs. total number of edits by each editor, for all 
editors (blue) and a group of ``Bad Editors'' with large number of penalties during their
editorial life (red). For both samples, the average value increases initially, meaning a trend towards 
more concentration on limited number of pages, followed by a decrease
signalizing broadening of filed of interest and editorial zone of editors as they get more experienced. 
The consenter at the peak is more intense for Bad Editors
and their contribution is focused on fewer pages compared to average editors.} \label{fig:edit-article}
\end{figure}

One source of the inhomogeneity is that statistical characteristics of editors show time evolution; different phases can be identified in the editorial behavior as they get more and more mature. For example, in Fig.~\ref{fig:edit-article} the number of edits per different articles
as a measure of the editors' focus,
versus total number of edits, calculated for a large sample of editors (all editors with at least 1 edit) 
is showed.
The general trend is as follows: Editors start by less intensive 
edits on different pages, then gradually they get 
more focused on few articles and
the ratio of the number of edits per number of different articles increases. Once they reach a level of maturity, 
again their field of interest becomes wider
and finally extremely senior editors, distribute they editorial efforts on a huge number of articles. 
Note that this is the average trend and it may differ from editor to editor. In the same plot, 
same quantity is shown for a sample of ``Bad Editors''
who have been blocked from editing at least 7 times per 1000 edits. Although the overall trend is the same as the whole sample, but a 
larger 
peak, indicating more intensive focus 
on few pages, is clearly visible. This is very intuitive: Editors seeking conflict and having the tendency to violate the guidelines have 
special interests in a limited number of pages, where they disturb the collaborative environment. 

\subsubsection{Time of editing}\label{sec:edit-time}
Since all edits are recorded along with a timestamp, it is very convenient to 
perform temporal analysis on editorial activities at different time scales.

\begin{figure} \sidecaption
\includegraphics[width=1.0\textwidth]{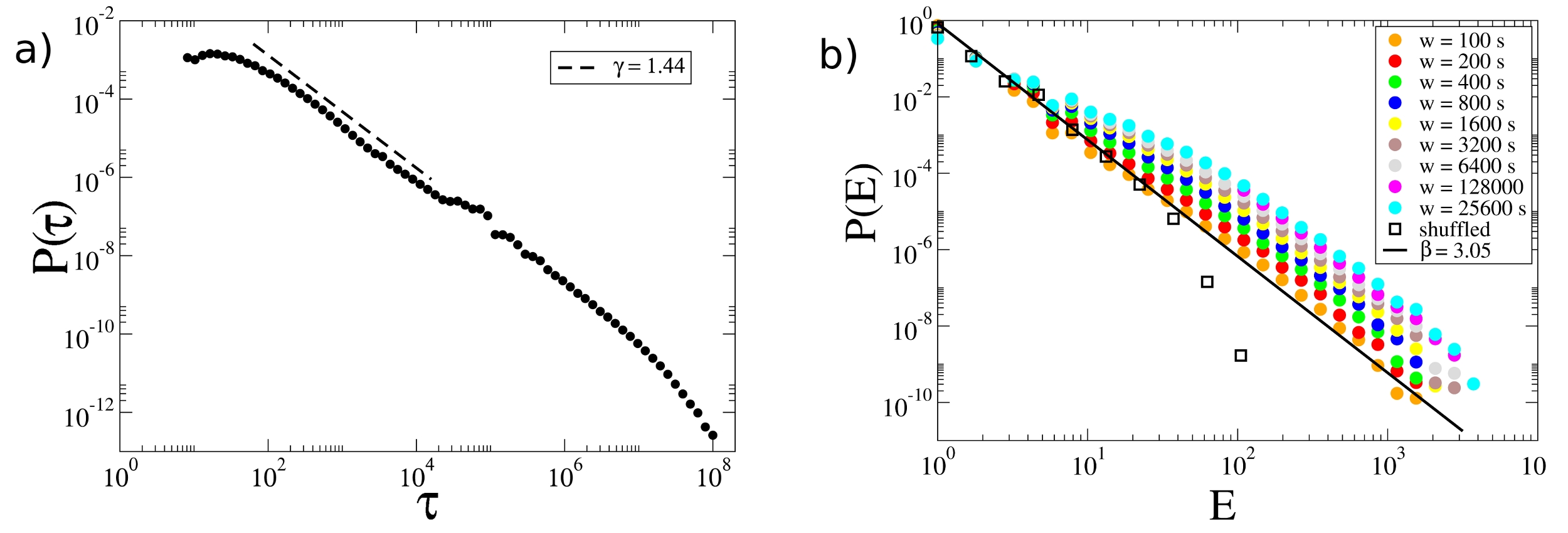} 
\caption{Two characterizing functions of the temporal pattern of editorial activity at 
individual editors level averaged over a sample of 100 most active editors. 
a)  Probability distribution function of the inter-edit time intervals (in seconds) fitted with a power 
law of the exponent $\gamma=1.44$. The bump is due to the circadian patter and corresponds to 24 hours. 
b) Probability distribution function of the number of edits in the bursty periods separated 
by windows of silence with the width of at least  $w$. Color circles are the original data
and empty squares corresponds to the shuffled sequence. The exponent of the power-law fit is $\beta=3.05$ and 
the decay for the shuffled data tends to an exponential form. From panels (b) 
it becomes clear that there is long term correlations in edit trains at the level of editors in 
addition to broad distribution of time intervals shown in panel (a). {\it These
figures were originally published in \cite{yasseri2012b} under the terms of the 
Creative Commons Attribution License.}} \label{fig:user-burst}
\end{figure}

\paragraph{Burstiness:} Most of the editors 
do their edits following a certain type of 
inhomogeneous temporal pattern. There are periods or bursts of high activity separated by low or no activity intervals. 
Compared to a homogeneous Poisson process
% with a fixed rate of events in time, distribution of time intervals between successive events 
the distribution of the inter-event times has a much fatter tail in the case of a bursty pattern. One trivial source of the temporal inhomogeneity is the circadian pattern of human activity. However,
recently it was shown by introducing the notion of bursty periods that ``burstiness'' often origins
from memory effects and reinforcement mechanisms \cite{karsai2012}. These bursty periods are separated by intervals of length $w$ of no activity and the distribution of the number of events in the bursty periods follows a power law in contrast to the memory free case, where it is exponential. Our investigations on short time scale temporal features of editorial activities, reveal
strong evidences for the presence of similar mechanisms \cite{yasseri2012b}. 
For instance, in Fig.~\ref{fig:user-burst} two characteristic measures of temporal 
patterns for the activity train of a class of active users are shown.
First, the distribution of inter-edit time intervals (Fig.~\ref{fig:user-burst}~(a)) 
is extremely fat-tailed, indicating the presence of long silence periods. Second, the distribution of the number of edits in bursty periods follows a power law, which is not sensitive to the choice of $w$  (Fig.~\ref{fig:user-burst}~(b)).
However, in the absence of memory effects, these the latter would be an exponential distribution
\cite{karsai2012}, 
as it is indeed the case when we shuffle the data
Fig.~\ref{fig:user-burst}~(b). 

Ung and Dalle, also reported a power law distribution of the inter-edits 
time intervals and interpreted their observation as an outcome of editors' focus
on few certain tasks (articles). They measured the slope for different 
class of users and showed that more/less skewed distributions correspond to
more focused/dispersed editors \cite{ung2010}.

\paragraph{Daily patterns:} 

As mentioned above, the activity pattern of individual editors are quite heterogeneous in time. 
However, if we consider the whole editorial pool of a language edition 
of Wikipedia, we can define an average activity
level for all editors which also has its own large scale characteristics. 
In \cite{yasseri2012a}, it is shown that WP is mostly edited between
 1~pm and 11~pm, almost in a universal manner for all 
language editions. This is in accord with the results in \cite{reinoso2011,karkulahti2012}. 
Deviations from this universality originates from cultural differences and 
working habits, such that language editions with more
editors from countries with longer working hours, are even more edited in 
later time in evening and around midnight.
In addition, for more global language editions, the activity curves are 
flattened, due to contributions from different time zones (see \ref{sec:origin}).

\paragraph{Weekly patterns:} 
%In contrast to daily patterns of editorial activity, 
Weekly patterns are quite universal within one WP and different WPs
can be classified in different categories based on the activity pattern of 
their editors \cite{yasseri2012a}. For example, 
German, English, Spanish, and Italian WPs are mostly edited during 
the working days, in contrast to Japanese, Korean, and Chinese WPs being mostly
edited on weekends. Our findings are in accord with \cite{reinoso2011} but in contrast with \cite{karkulahti2012}. However, the latter work studied a sample of four languages only and a shorter monitoring time, and we believe that these lead to the conclusion that editorial activity in WP ``while showing a clear diurnal pattern, 
do not have a clear weekday-weekend pattern.''

\subsubsection{Edits origin} \label{sec:origin}
As mentioned earlier in Section.~\ref{sec:editors}, personal information of editors is rarely available. 
That includes their nationality and living place. However, to understand many
aspects of social characteristics of the editors societies, as well as conflicts and potential biases 
in content, such information could be crucial.
To achieve exact data on the location of editors, analysis must be restricted to 
unregistered users with edits recorded along with IP addresses, whose edits are typically  between 5\% to 10\% of the
total community contributions in different language editions, and clearly not representing the whole community. Moreover, a considerable part of such editors are atypical (vandals, single act editors). Nevertheless,
Hardy et~al. followed edits of 2.8 Million such editors and geolocated them and the edited articles. 
By counting the number of edits as a function of distance between
editor and article, an exponentially decaying distance dependence was 
%gravitational model is 
obtained \cite{hardy2012}.
Cohen has investigated the contribution of unregistered editors to English WP and concluded 
that most of unregistered edits are from large cities and metropolitan areas
\cite{cohen2010}. However, normalization to population of regions 
seems to be a missing essential for such conclusion.

\begin{figure} \sidecaption
\includegraphics[width=1.0\textwidth]{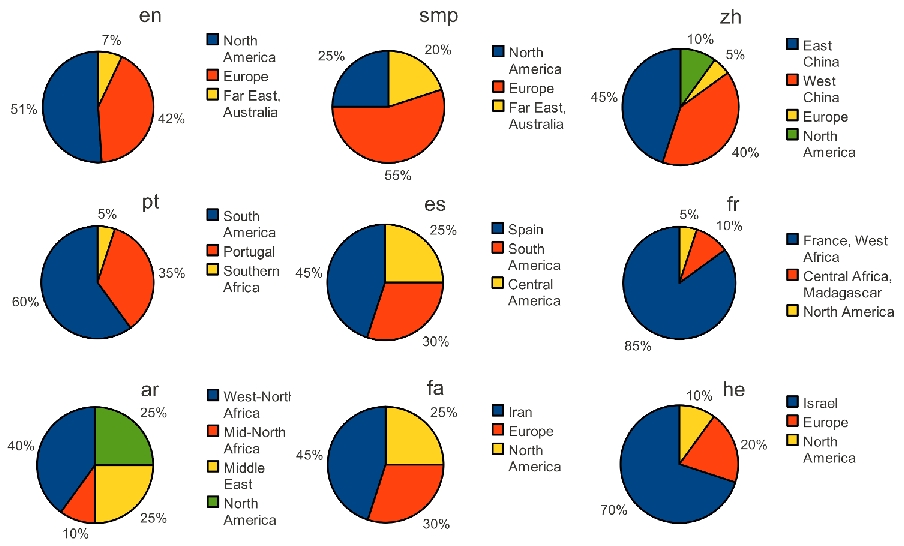} 
\caption{Estimations for the share of editors from different regions to different Wikipedia
language editions. In the first row, {\it en}, {\it smp} and {\it zh} stand for English,
Simple English and Chinese WPs. The share of North America to the English WP 
hardly goes beyond half and it is around one quarter for Simple English WP.
In the middle row, {\it pt}, {\it es}, and {\it fr} stand for Portuguese, Spanish, and French. The low level of 
contributions to French WP from North America (Canada) is worth mentioning.
In the lower row, three Middle East languages are shown; {\it ar}, {\it fa}, and {\it he} stand for Arabic, 
Persian, and Hebrew. Here, large amount of contributions from western regions to Persian WP
is notable. {\it This 
figure is originally published in \cite{yasseri2012a} under the terms of the 
Creative Commons Attribution License.}}   \label{fig:share}
\end{figure}

Based on the results on daily patterns of editing for geographically 
localized WPs, such as German, Italian, Hungarian and defining a ``standard activity pattern'', 
one could estimate the global
distribution of editors to global language WPs in the following way; initially some 
candidate regions, from which large population of editors would
contribute to the given WP are selected. In the next step, a linearly weighted 
superposition of standard activity patterns shifted to the local time of the candidate regions
is made. By minimizing the difference between this composed activity pattern and the activity 
pattern constructed from the real data, a set of optimal weights is obtained.
Clearly, these weights are proportional to the share of editors contributing from the corresponding 
regions. Surprisingly, 
it turned out that English WP is almost equally edited by North Americans
and editors from the rest of the world \cite{yasseri2012a}. In Fig.~\ref{fig:share}, 
estimations for the share of contributions from different regions to each 
language edition are shown. 

\subsubsection{Characterization of edits}
As mentioned above, each editor has her own unique characteristics and editorial personality. 
However, similar patterns could be observed by considering types of 
edits. In a novel approach, Wettenberg et~al. established a visualization 
method to illustrate different editorial actions, e.g. adding,
spelling correction, reverting, etc. in a time sequence. In the next step based on the patterns of 
activity they could distinguish different kinds, namely
systematic activity, reactive and mixture activity patterns \cite{wattenberg2007}.

Kittur et~al. have classified editors based on number of edits and also specifically followed admins' 
contributions from the inception of WP \cite{kittur2007}.
They concluded that in the beginning of the WP history, large amount of contributions were 
offered by ``elite users, however, it has gradually changed in a way that
after 2004, average users overtook the elites. By counting the number of added and removed words 
for different editors, they suggest that elite users in average add more words per
edit compared to normal editors.

\begin{figure} \sidecaption
\includegraphics[width=0.6\textwidth]{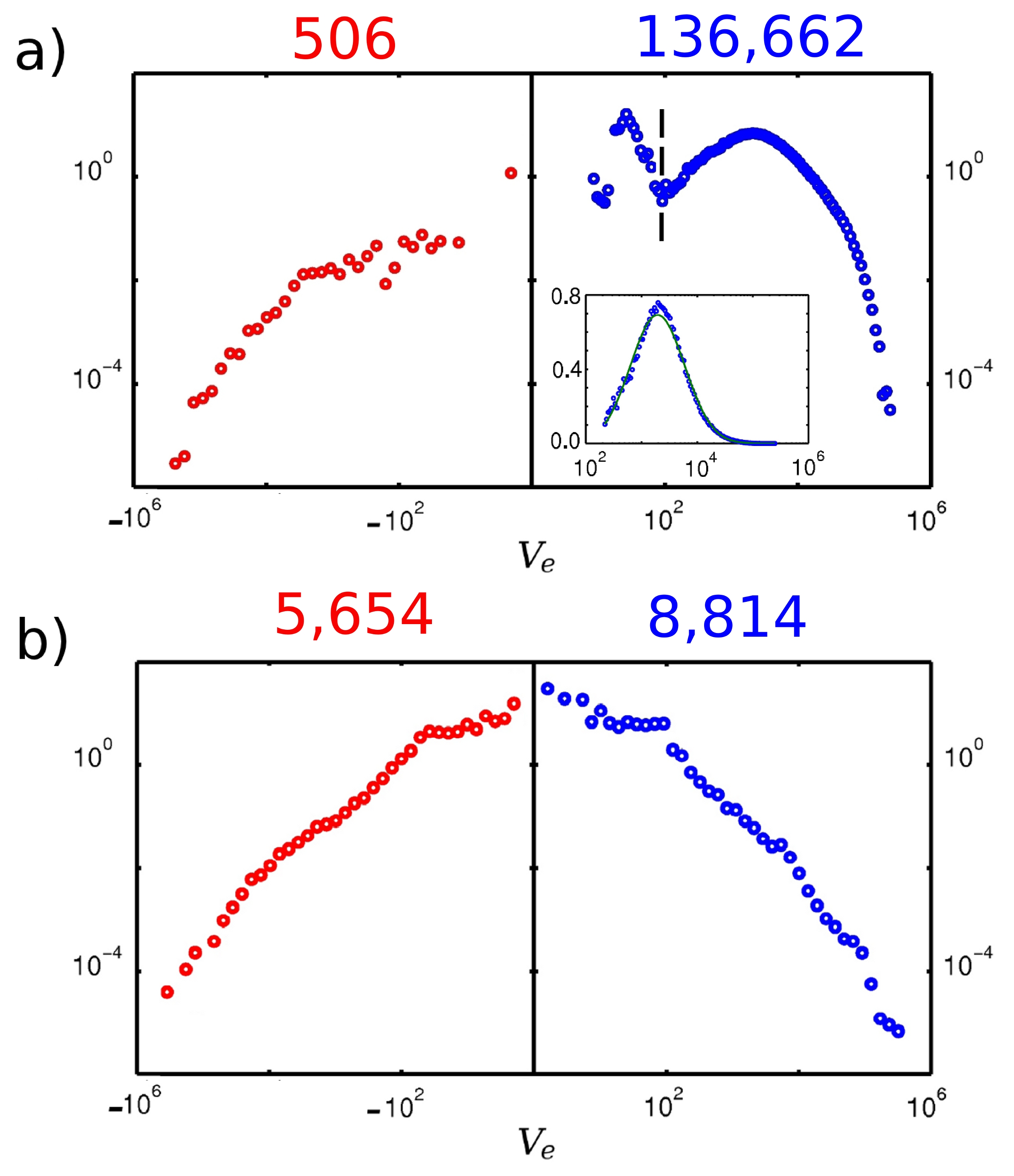} 
\caption{Probability distribution function of edit volume (in bytes) for two 
typical editors: producer (a) and maintainer (b). 
Blue/red points representing the volume of added/deleted words in each edit. Total 
number of edits of each type are reported on top of each panel. The producer editor
has much more adding edits, whereas the maintainer has comparable number of adding, 
deleting edits. The distribution of the volume of added parts by the producer editor
has two peaks corresponding to volumes of few sentences and few paragraphs respectively 
(including the references and Wikimedia tags). {\it Inset} of panel (a) is
a semi-logarithmic zoom on the right part of the added words volume distribution separated by the dashed 
line in the main panel. Fitted line to the inset is a log-normal distribution function.} \label{fig:volume}
\end{figure}

We considered the volume of contributions by measuring the volume of each edit in unit of characters. 
Naturally, negative volume is assigned to deletion. 
In Fig.~\ref{fig:volume} two examples of edit volume profile of two different type of users is shown: 
Fig.~\ref{fig:volume}~(a) shows a typical ``producer'' 
and Fig.~\ref{fig:volume}~(b) shows a typical ``maintainer'',
It is needless to mention that there are 
many different type of users, but by analyzing the edit volume profile of 
few tens of most active editors, these two types are found dominantly. In other words, although the results shown here are for two single editors, they are representative of large groups of editors, who can be categorized in one of the these groups based on their edit volume profile. Note that, for the case of producer editors, a separation between additions in the size of 
``sentences (first peak) or ``paragraphs'' (second peak) is nicely visible.  

%%%%%%%%%%%%%%%%%%%%%%%%%%%%%%%%%%%%%%%%%%%%%%%%%%%%%%%%%%%%%%%%%%%%%%%%%%%%%%%%%%%%%%%%%%%%%%%%
\subsubsection{Linguistic features}\label{sec:language}
The content of Wikipedia is generated by large number of editors 
collaboratively and without any professional or external supervision.
That makes the resulting written language of WP articles a 
unique multilingual corpus of natural languages.
A single sentence in WP might be written, edited and polished 
by various editors many times, therefore any personalization
bias is eliminated on large scales. Moreover, the fully recorded history 
of articles give the opportunity to follow the
short time scale evolution of language and characterize the gradual changes 
of written language in the digital era.
Finally, since WP is huge, and available in many different languages,
statistical approaches 
%in ``thermodynamic limit''
can be taken in a proper way. 

In a practical perspective, Tyers and Pienaar used WP to extract 
pairs of corresponding words in different languages \cite{tyers2008}.
Serrano et al, used WP corpus along with two others, to build up  
statistical models concerning fundamental concepts of patterns of word 
appearance in the text and
 vocabulary size \cite{serrano2009}. Gabrilovich and Markovitch, introduced 
a method to calculate semantic relatedness of text fragments by extracting a ``high-dimensional
space of concepts'' from WP \cite{gabrilovich2009}.  In a recent paper \cite{kornai} Kornai argued that the maturity of WP and the activity on it are 
important indicators for the chances of survival of a language in the digital age.

Our approach to WP as a text corpus is based on readability measures. 
We analyzed the readability of the English WP by applying the "fog index" \cite{gunning1952,gunning1969}, a simple empirical 
formula suggested by
Gunning in the middle of the last century for the English language:
\begin{equation}
F= 0.4(\frac{\textrm{\# of words}}{\textrm{\# of sentences}}+100\frac{\textrm{\# of complex words}}{\textrm{\# of words}}).
\label{eq:gunning}
\end{equation}
where, complex words are those with three or larger syllables. 
The readability measure $F$ is interpreted as the 
length of needed education time in years, to be able to
read and understand the text.

We found out that the overall $F$ of English WP is 
high with $F=15.8 \pm 0.4$ compared to other standard English corpora, for instance
British National Corpus\footnote{\url{http://www.natcorp.ox.ac.uk}} with $F=12.1 \pm 0.5$ \cite{yasseri2012c}. 
However, readability is not homogeneous among articles in different topics.
We observed that articles on more sophisticated topics or concepts, especially in science and 
philosophy are less readable than, e.g., biographical articles.

An interesting language edition of WP is ``Simple Wikipedia'', which 
is meant to be a proper reference for readers with weaker knowledge of 
English, e.g.,
children, language learners or non-native speakers. Editors of Simple WP are explicitly 
requested to use a simpler language, limited vocabulary, less complex words, shorter sentences, and easy
structures.\footnote{\url{http://simple.wikipedia.org/wiki/Wikipedia:How_to_write_Simple_English_pages}}

In a recent work \cite{besten2008}, Simple is examined by measuring the Flesch reading score
\cite{flesch1979} and it is  found that Simple is not simple enough compared to
other English texts, however with a positive trend in time towards more simplicity. There 
have been also attempts to use Simple WP for establishing
text simplification algorithms \cite{yatskar2010,napoles2010,coster2011}, 
however with the assumption that Simple WP is really simple.
The comparison of Simple and English WPs enables to study the ability of the editors to 
fulfill a preset task (namely enhance readability) and, at the same time, it also sheds more light to the concept of language 
complexity in general. We measured the readability index for a sample 
extracted from Simple WP \cite{yasseri2012c}. We fund it to be $10.8 \pm 0.2$, i.e., indeed much lower than for the English WP but just
as large as a corpus made of Wall Street Journal\footnote{\url{http://www.wsj.com}} articles. 

To further analyze language complexity of Simple WP, we made the statistics 
at the word level, and surprisingly observed that vocabulary richness of Simple
is comparable to that of main English WP. Moreover, by examining two fundamental 
laws of linguistics, namely Zipf's law \cite{zipf1935}
and Herdan-Heaps' law \cite{Herdan:1964,heaps1978}, we again confirmed that 
vocabulary richness of Simple and Main English WP are not 
significantly different \cite{yasseri2012c}, although the directives explicitly suggest self-restriction in this respect for Simple editors. 
Detailed analysis of longer units (n-grams) of words  shows that the language of Simple is indeed less complex than that of Main but 
due to more frequent use of predefined language blocks, e.g., chains of words in the length of 4 or 5 words in Simple. 
Lengths of sentences are also shorter in Simple compared to Main.
One can conclude that Simple editors solved fairly well the task to write more readable 
texts as compared to those in the main English WP without following slavishly the directives but mostly by reducing the variation of language compounds.

%%%%%%%%%%%%%%%%%%%%%%%%%%%%%%%%%%%%%%%%%%%%%%%%%%%%%%%%%%%%%%%%%%%%%%%%%%%%%%%%%%%%%%%%%%%%%%%%
\subsection{Conflicts and edit wars}\label{sec:conflict}

In the process of creating 
a common product by various agents, occurrence of controversies
due to different opinions are unavoidable. WP is 
neither an exception
in this sense. WP editorial wars and disputes are known and studied 
phenomena \cite{kittur2007,brandes2008,vuong2008,apic2011,sumi2011a,sumi2011b,yasseri2012b}. Editorial wars could be evoked both by internal and external causes. 
For example life events of celebrities \cite{yasseri2012b} 
or natural disasters \cite{keegan2011} could conduct flows of editors to
 an article leading to tensions and disagreements. 
Apic et~al. showed that disputes in WP are corresponding to real
 world geopolitical instabilities in many cases \cite{apic2011}. 
To study editorial wars in details, the first step
is to establish an algorithm to locate and rank the debated articles among 
the relatively large number \cite{sumi2011a} of peacefully written ones.
There are different proposals for this goal in the literature \cite{kittur2007,brandes2008,vuong2008,sumi2011b}.
In the following Section we briefly describe our previously established 
method \cite{sumi2011b} for locating and ranking editorial wars.

\subsubsection{Identification of controversial articles}
\begin{figure}\sidecaption
\includegraphics[width=0.5\textwidth]{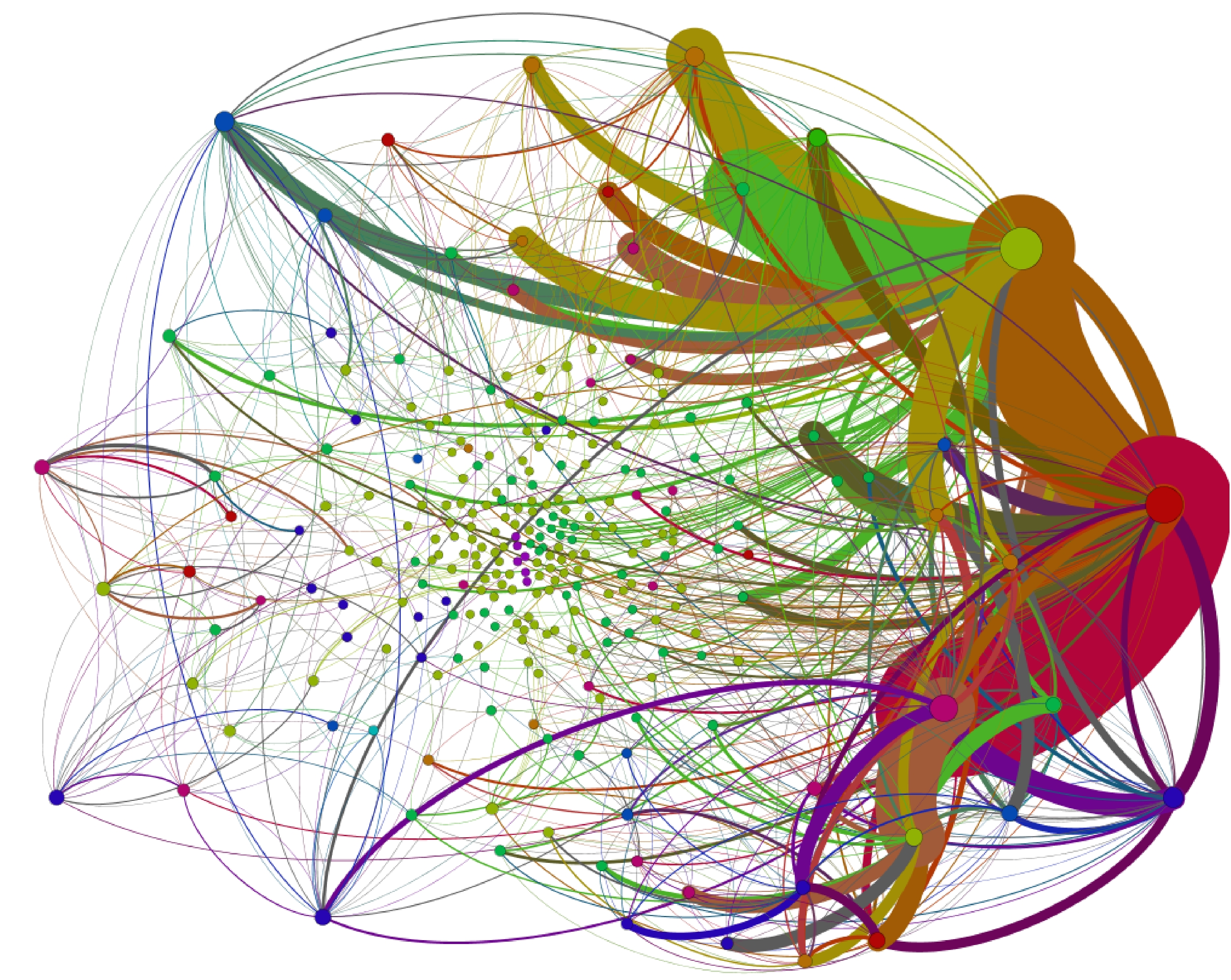} 
\caption{A network representation of reverts in the history of the article on ``Anarchism'' 
in English WP. Nodes are editors and links are representing reverts.
Size of the nodes is proportional to the total number of reverts, in which the editor is 
involved and width of the links is proportional to the number of total reverts between 
their corresponding pair of editors. Few nodes are strongly connected whereas most of the nodes 
are connected only through weak links. Completed triangles 
are clearly under-represented. {\it The graph is generated Gephi, 
the open graph visualization platform, available at http://gephi.org}. } \label{fig:revert}
\end{figure}

The algorithm to identify disputed articles, introduced in 
\cite{sumi2011a,sumi2011b,yasseri2012b}, is based on counting mutual reverts
by pairs of editors, i.e. when an editor undoes another editor's edit
completely. To detect reverts, we first assign a MD5 hash code \cite{MD5} to
each revision of the article and then by comparing the hash codes, detect when
two versions in the history line are exactly the same. In this case, the
latest edit (leading to the second identical revision) is marked as a revert,
and a pair of editors, namely a reverting and a reverted one, are
labeled. A ``mutual revert'' is recognized if a pair of editors $(x,y)$ is
observed once with $x$ and once with $y$ as the reverter. The weight of an editor
$x$ is defined as the number of edits $N$ performed by her, and the weight of a
mutually reverting pair is defined as the minimum of the weights of the two
editors. The controversiality $M$ of an article is defined by summing the 
weights of all mutually reverting editor pairs, excluding the topmost pair, 
and multiplying this number by the total number of editors $E$ involved in 
the article. In formula,
$$
M = E \sum_{\text{all mutual reverts}} min(N^{\rm d}, N^{\rm r}),
$$
where $N^{\rm r/d}$ is the number of edits on the article committed by
reverting/reverted editor. The sum is taken over mutual reverts rather than
single reverts because reverting is very much part of the normal workflow,
especially for defending articles from vandalism. The minimum of the two
weights is used because conflicts between two senior editors contributing more
to controversiality than conflicts between a junior and a senior editor, or
between two junior editors. In Fig.~\ref{fig:revert}, a network representation of mutual reverts of the article 
on ``Anarchism'' in English WP is shown. 

Clearly, as 
the time goes on, more mutual reverts could happen in the history of
the article. This makes $M$ a dynamic, monotonically increasing variable. 
Having calculated $M$ for all articles, we are able to find and rank most disputed ones and
investigate them in details. We carried out a detailed comparative study of possible single measures 
and found that $M$ is in most cases as good as its alternatives if not better with the additional advantage of being applicable to different languages. 
The superiority of our single parameter measure was reinforced by a recent independent investigation \cite{hoda2012a}.

\paragraph{Controversial topics:} Based on the calculated 
controversy measure for articles in different languages, 
first conclusion is that, although there are sever editorial wars on some articles, 
but most of the articles in different languages evolve
rather peacefully.
However, the 
truly disputed articles consume a considerable amount of editorial resources. 
Interesting patterns are observed by comparing the debated titles in different language editions
\cite{yasseri2012d}. For instance, issues related to politics and religion 
are 
commonly 
among the most disputed articles
in many language WPs, whereas, some category of topics only become controversial in 
specific languages. Science and philosophy in French 
and soccer clubs in Spanish WPs are examples of locally debated 
topics. There are even articles, in top of the controversy list in one
WP, which is not even covered in other language edition, or 
does not have a separate article. Here examples are detailed
articles around ``Baha'i Faith'' related topics in Persian WP. Finally, 
surprisingly, in the Hebrew WP,
sport is debated as much as religion and politics.

\subsubsection{Temporal features}
The understanding of the emergence of conflicts, 
their escalation and resolution is important for maintaining 
WP and may give hints in general for techniques of conflict management. The controversiality measure $M$ enables the temporal analysis of  
editorial wars on short and long time scales. 

\paragraph{Edit frequency:} Intuitively more popular articles are subject to 
more collision of opinions and edit wars. However, 
the correlation between the average times between edits and the measure $M$ is not significantly strong ($C=-0.03$).  

\begin{figure}\sidecaption
\includegraphics[width=0.75\textwidth]{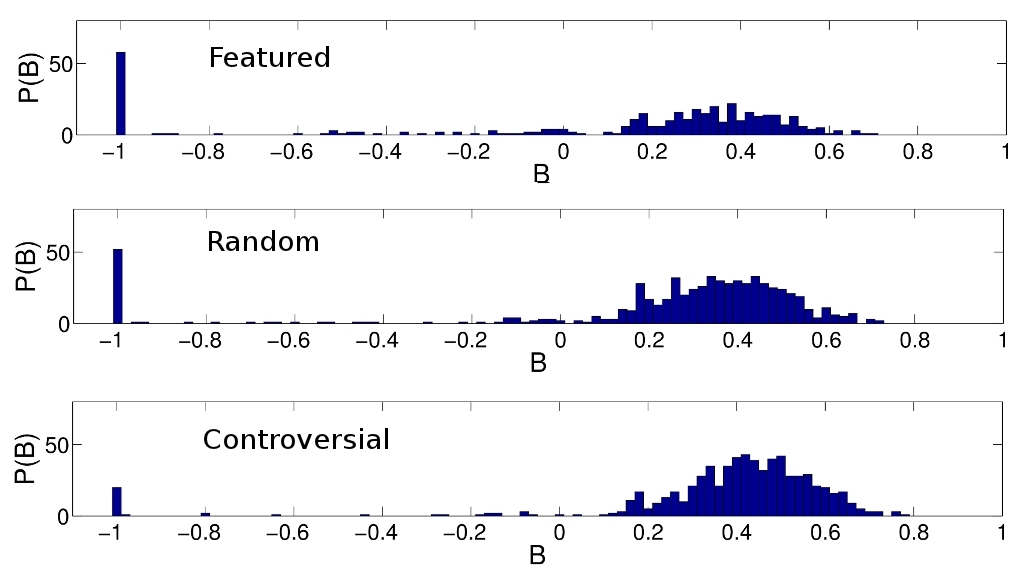} 
\caption{Histogram of $B$ values calculated only considering the temporal sequence of mutual reverts, for 3 different samples of articles; randomly selected, featured, and controversial articles.
The last two samples were made based on the lists
provided in Wikipedia (see Section~\ref{sec:article}).} \label{fig:b}
\end{figure}

\paragraph{Burstiness and conflict} As discussed in 
Section~\ref{sec:edit-time}, bursty trains of activity
are clearly present on the editor level. Now, we focus on the edit trains of 
individual articles, i.e., for a particular article we consider the edits from all editors. We calculate the burstiness measure
suggested in \cite{goh2008} as
\begin{equation}
 B \equiv \frac{\sigma_{\tau} - m_{\tau}}{\sigma_{\tau} + m_{\tau}},
\end{equation}
where $\sigma_{\tau}$ and  $m_{\tau}$ are the standard deviation and 
the average of inter-edit time intervals of each article. 
For a regular pattern with a delta function-like interval 
distribution, $B \rightarrow -1$, and for a random Poissonian process
with a fixed event rate $B \rightarrow 0$. However, for fat-tailed 
distributions of time intervals, when the standard deviation diverged for
infinite systems,  $B \rightarrow 1$. 
The correlation between burstiness and controversy was also found to be  
rather small (C=0.05), considering the whole sample of articles in English WP. We also considered smaller 
samples of ``featured'' and ``controversial'' articles, based on the lists described in Section~\ref{sec:article} and compared them 
to a sample of articles selected randomly. When looking at the distribution of the burstiness $B$ we could not find significant 
differences between the three categories of articles. However, the $B$-values for the reverts, and especially for mutual reverts show 
significant increase of the burstiness, when going from featured through average to conflict articles (see Fig.~\ref{fig:b}).

\begin{figure}
\includegraphics[width=1\textwidth]{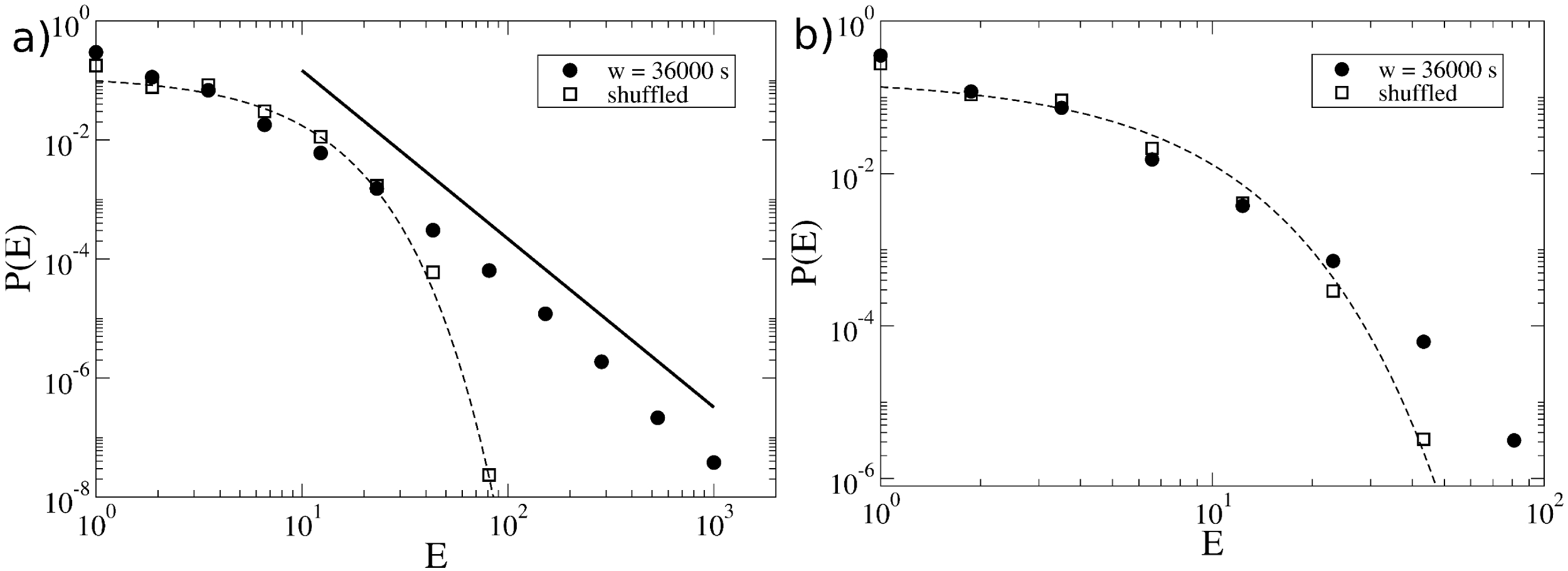} 
\caption{Probability density function of the number of edits in bursty periods separated by a silence window of 10 hrs, 
for two samples of (a) highly controversial, and (b) peaceful 
articles with the same average inter-edit time interval. Black circles are the original data and 
empty squares are the shuffled sequence of the same intervals. 
Bursty periods with very large number of events are visible in (a), whereas the decay
of the probability density function is very close to exponential and indeed to the density function for the shuffled data in (b), indicating presence of memory effects in 
the case of controversial articles. 
{\it This figure is originally published in \cite{yasseri2012b} under the terms of the 
Creative Commons Attribution License.}} \label{fig:article-E}
\end{figure}

To further investigate the temporal features of the editorial wars, 
we select two small samples of articles
with the same average inter-edit time intervals, however one collected 
from largely disputed articles, and the other consisting of 
peaceful ones. Fig.~\ref{fig:article-E} clearly shows that 
in the sample of disputed article, edits come in condense bursty trains
with long range memories, whereas the statistics of the number of edits in the bursty periods in peaceful articles is much 
closer to an uncorrelated process (signalized both by the data point from the shuffled sequence and the exponential fit).
In a recent paper
\cite{kampf2012}, it was claimed that edit time series can be described by a Poison process, i.e., 
``edit-events are only short-term correlated''. What we can deduce from our data is that the inter-event time distribution is much less fat tailed for peaceful articles than for conflict ones.
Note that due to the overwhelming number of peaceful articles 
a random mixture of articles does not sample conflicting ones. Therefore the conclusion of \cite{kampf2012} is true at most for such articles but, as we demonstrated, 
not for controversial ones, where long-term memory is definitely present.

\subsubsection{Talk pages, conflict and coordination}
As mentioned in Section~\ref{sec:access}, talk pages are  channels to resolve 
the editorial disagreements in a more civilized manner than
overriding each others edits and ``talk before you type'' is considered 
as the ideal mechanism of coordination in WP \cite{viegas2007}.
In a novel approach, Hautasaari and Ishida investigated the role of talk pages 
in coordination of translation of articles from English to Finnish, French, and Japanese 
\cite{hautasaari2012}. They conclude that 
most of the debates in this field are about naming issues and not much about the content.
Schneider et~al. performed very detailed analysis of Talk
pages from 100 articles manually and talk pages
from 5000 articles quantitatively \cite{schneider2010}. 
Their results for the category of controversial articles suggests ``significant 
variance between discussion threads (different sub-topics in the talk page of a certain article) on their talk
pages'', such that the distribution of the length of single threads is quite heterogeneous. 
Many threads are rather short, with few comments, and few of
them become extremely long with numerous comments. This is in accordance with the results of
\cite{gomez2011}, where a preferential attachment model
to explain the discussion cascades in the talk pages was presented.

We measure the length of the talk pages for all articles. 
The correlation between talk page length and
$M$ for the English WP is much more significant (C=0.54) than that with the edit frequency. 
It indicates that most of the debates are 
reflected in talk pages simultaneous to edit wars directly on the article. This is 
partially supported in \cite{kaltenbrunner2012}, where a method to detect
``peaks'' in talk pages is presented and showed that larger peaks mostly 
co-occur with peaks in editorial activities in a distance of 2 days.
However, there are substantial differences between WPs on different languages in the usage of talk pages. In general, less developed WPs use talk pages less but even rather mature WPs, like the German one do not fight out controversies on the talk pages. (For a collection of visualizations and other related materials to edit wars, 
see \url{http://wwm.phy.bme.hu/}.)

\paragraph{Discussion Networks:}
In contrast to the revert network of editors, which can be constructed rather 
straightforwardly, creation of 
talk page networks need more sophisticated algorithms. Laniado et~al. constructed 
three types of talk networks by considering i) direct replies between users
in article discussion pages, ii) direct replies in user talk pages, and iii) personal 
messages posted on the talk page of another user \cite{laniado2011}.
The conclusion of this studies suggests the presence of dissortativity in outgoing links 
and assortativity in incoming links. 

Our case study of the talk page of ``Safavid dynasty'' showed that 
most of the comments are exchanged between few editors, 
who are actively editing the articles. In addition, the occurrence of 
clusters is very rare, such that most of the conversations
are between pairs of editors and not bigger groups of them. 
This is in accord to analysis on mutual reverts which shows only few editors
are responsible for large amount of edit wars \cite{yasseri2012b}.
By fine investigation on those few user-names very active in controversial articles, 
we could recall many of them from our list of ``Bad Editors''
introduced in Section~\ref{sec:stat}. 
\paragraph{Language complexity and sentiment:} Laniado et al, studied the emotional 
aspects of talk page discussions by measuring sentiment of the comments and found
``replies are on average more positive than the comments they reply to, and editors having 
similar emotional styles are more likely to interact with each other.''
Moreover, they found that editors with more social power, i.e. admins, talk more positively 
and interestingly this is also the case for female editors \cite{laniado2012}.

We measured the readability of talk pages based on Eq.~\ref{eq:gunning} and compared it to the 
readability of articles for two samples of controversial
and peaceful articles. In both cases there is a significant reduction in readability, going 
from articles to corresponding talk pages \cite{yasseri2012c}. However, 
the reduction is much more significant for the controversial articles. This can be explained 
by previous sociological theories on the effect of destructive conflict
on complexity reduction of language \cite{samson2010}; In simple words, when people talk with 
more temper, they use less sophisticated language.

\subsubsection{Leader-follower behavior in conflict}
The community of editors is structures though it is not easy to unfold its patterns. When studying the talk pages 
of highly edited articles, it becomes clear 
that editorial behavior is influenced beyond the content also by personal relationships \cite{rung}. 
There are dominant editors and others, 
who only follow them. Such relationships largely influence the emerging editor network. The easiest way to detect related behavior 
is to concentrate on leader-follower pairs. These are pairs of editors (say, $A$ and $B$), who often act in a 
specific order, i.e., $A$ always precedes $B$ within a reasonable time, e.g., 1 day. As we are interested in 
the difference between peaceful and conflict articles, we concentrated on the leader-follower phenomenon in 
reverts \cite{rung}. We defined the following process as an event: $A$ reverts $C$ and (within one day) $B$ reverts $C$, 
where $C$ is fixed only for this specific event. Confining our interest to reverts restricted considerably 
the statistics, however, significant differences between the two groups of articles could be observed here. 

We took two different edit history samples of WP. The first sample consisted all reverts of the 837 
articles with $M$ value above $10^5$ (conflict articles) ordered by time. In order to avoid the effects of vandalism, we 
excluded reverter-reverted editor pairs consisting at least of one IP address or bot. Moreover, to gain a better focus on leader-follower relationships and not the 
effects produced solely by editorial wars between two editors, we also excluded repeated 
reverts where the reverter-reverted pair was the same and no other reverts happened between these two 
reverts. This seed consisted of 303397 reverts. We took a sample of 12470 
articles with $M$ value under 500 (peaceful articles), where the number of reverts was approximately the same. We also created randomized versions of these samples.

\begin{figure}\sidecaption
\includegraphics[width=.75\textwidth]{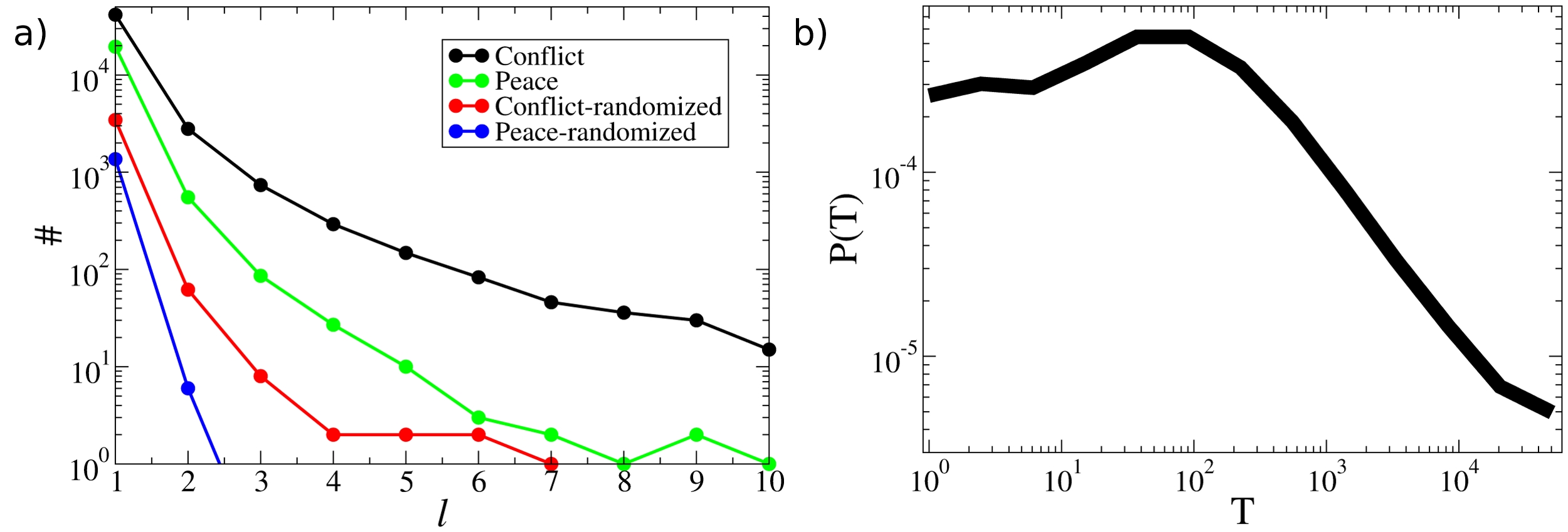} 
\caption{Leader-follower statistics from revert series of peaceful and conflict articles as well as their randomized versions. 
a) Statistics of leader-follower sequences. b) Probability distribution function of elapsed time between leader and follower events.} \label{fig:L-F}
\end{figure}

The results are summarized in Fig.~\ref{fig:L-F}~(a), from which it is clear that conflict articles have an 
enhanced amount of leader-follower patterns. This is expected as in this case parties are formed, where 
the hierarchy of editors can manifest itself. We see that even sequences of length $l=10$ occur. Fig.~\ref{fig:L-F}~(b) 
indicates that leader-follower actions have a characteristic time $T$ of about 2 minutes. This is a surprisingly short 
period and underlines the personal rather than contextual motivation. 

\subsubsection{War scenarios}\label{sec:warscenario}
The characterization of the temporal evolution of conflicts is crucial for their typology and understanding. Our measure $M$ is particularly suitable for such a study. 
We investigated controversial articles of English WP from this point of view.
Instead of the real time we use the number of edits as a control parameter. This way we eliminate several sources of temporal inhomogeneities like maturing the whole WP, differences in the sizes of the articles, 
and external events motivating editors to focus on an article \cite{ratkiewicz2010b,ratkiewicz2010c}.

\begin{figure}
\includegraphics[width=0.8\textwidth]{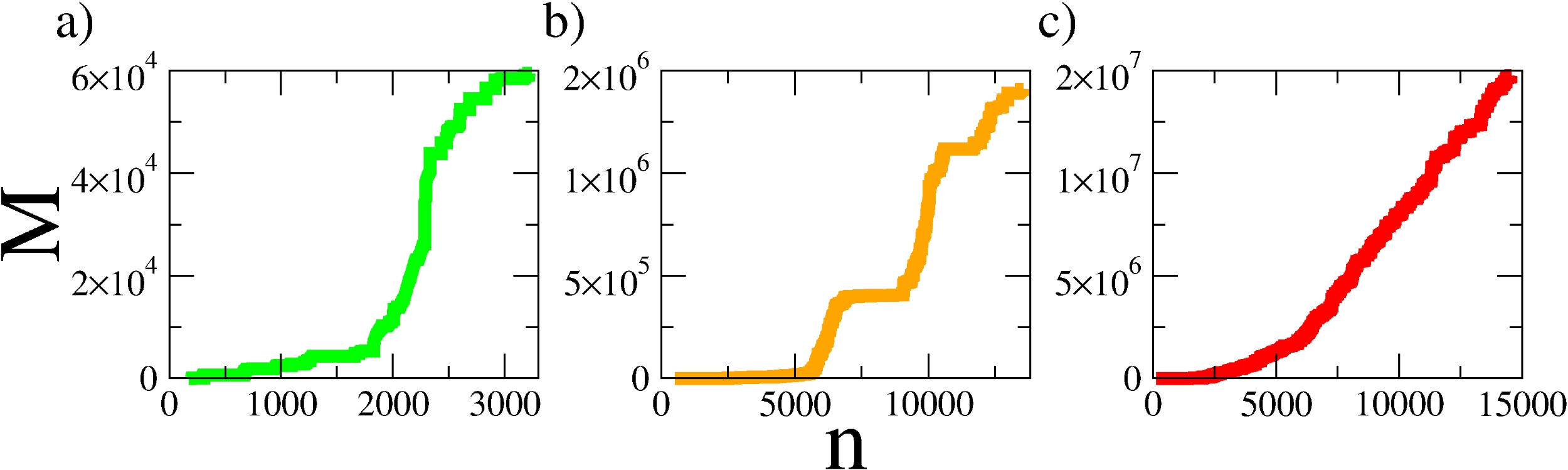} 
\caption{Evolution of the controversy measure $M$ as a function of number of edits on the articles $n$, for 
a) ``Bombing of Dresden in World War II'', b) ``Japan'', and c) ``Anarchism''. In (a), after a single 
period of editorial war, 
the article reaches the stable consensus, whereas in
(b) temporary consensus states are altered by new conflict periods. In (c) the rate of initiation of 
new conflicts is very high, such that no consensuses is ever reached. } \label{fig:M-n}
\end{figure}

Three different scenarios of wars could be distinguished \cite{yasseri2012b} from the temporal evolution of $M$ (Fig.~\ref{fig:M-n}): 

i) 
Consensus after war, Fig.~\ref{fig:M-n}~(a); 
After a smooth initial increase of $M$, an intense period of war appears and once
the conflict is resolved, the article reaches consensus and farther edits are mostly on
 polishing and improving the presentation quality.
This is the scenario for most of the disputed articles in English WP \cite{yasseri2012b}. 
ii) Stepwise conflicts, Fig.~\ref{fig:M-n}~(b); After the first cycle of conflict-resolution, the consensus
state might be altered mainly because of one of two reasons, namely occurrence of an external event which 
generates new controversy
or arrival of new editors, who are not satisfied with the previously compromised content of the article. 
Therefore, other conflict-resolution cycles may appear in the overall history
of the article. 
iii) Never-ending war, Fig.~\ref{fig:M-n}~(c); If the rate of incoming editors or 
external events related to the topic of the article, is considerably larger than the typical time to reach consensus, even a temporary 
equilibrium cannot be
achieved and the increase of $M$ becomes permanent. This is the case of highly popular and live-object articles. 
Number of such articles in English WP
does not exceed few hundreds (compared to some millions, the total number of articles).

\subsubsection{Agent-based modeling}
Motivated by empirical results on editorial wars in WP,
we aimed at providing a minimalistic agent-based model capturing the main features of the wars ~\cite{torok2012}. The model belongs to
the class of bounded confidence models of opinion dynamics introduced by Deffuant et~al. \cite{deffuant2000}.
It consists of two types of
elements; $N_{\rm{e}}$ editors and one article. In each Monte Carlo step, editors interact if their 
scalar opinions $x_i \in [0,1]$, $i=1\ldots N_{\rm{e}}$ 
are already closer to each other than a threshold value $\epsilon_{\rm{T}}$ and then they adopt the opinion of the arithmetic mean.
An editor edits 
the article if she finds it in a state $A \in [0,1]$ with a difference larger 
than $\epsilon_{\rm{A}}$ to $x_i$, otherwise she revises her own opinion which gets closer to the article state by an amount controlled by a parameter $\mu_{\rm{A}}$.
In addition, editors can be replaced in each step by new ones with a constant rate $p_{\rm{new}}$.

\paragraph{Fixed editorial pool:}
To evaluate the outcome of the model, initially $p_{\rm{new}}$ 
is set to 0, which leads to consensus for the whole parameter
space, meaning that after sufficiently time $A$ becomes constant. However, the relaxation time to consensus very much depends on the parameters set. 
There are three different scenarios to approach the consensus state:
i) for small values of $\mu_{\rm{A}}$, system needs astronomically long time to reach 
the final state, although $A$ is always very close to
the system average of $x_i$.  ii) Intermediate values of $\mu_{\rm{A}}$ puts the system into an 
oscillatory phase, in which $A$ fluctuates largely between two 
extreme values, however ending up with one of them in a relatively shorter time. iii) Large 
values of $\mu_{\rm{A}}$ leads to exaggerated fluctuations of $A$, however
with fast convergence of extremist editors and a shorter relaxation time compared to the previous cases.

\begin{figure}\sidecaption
\includegraphics[width=0.6\textwidth]{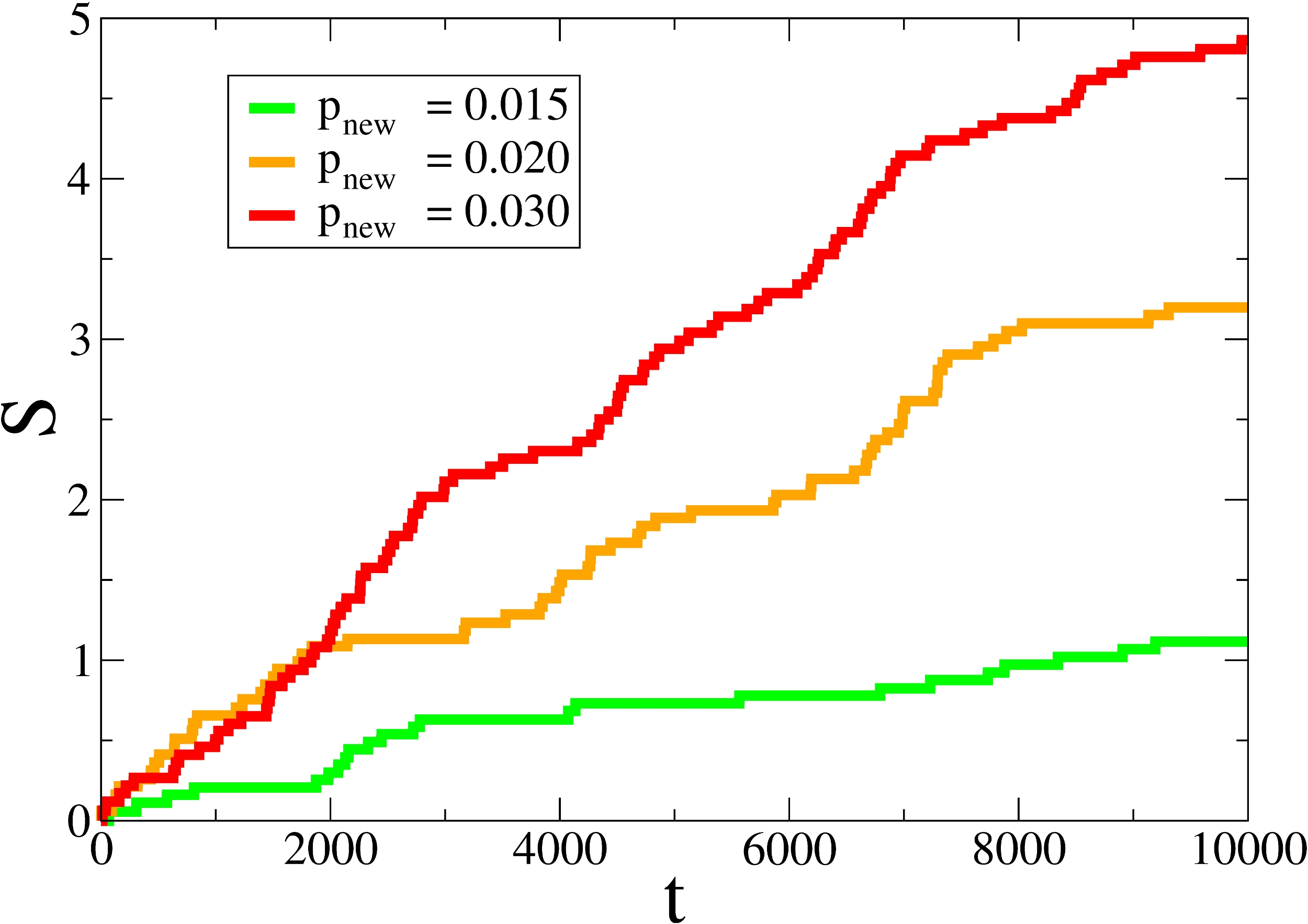} 
\caption{Evolution of the model analog to the controversy measure, $S$ as a function of number of edits on the articles $n$, for different rates of editor
renewal. Increasing the rate, the periods of peace get shorter and shorter, until they vanish entirely.} \label{fig:S-n}
\end{figure}

\paragraph{Dynamic editorial pool:}
The constant rate of replacement of old editors by new ones with random opinions 
can hinder the system to reach a time-independent state. 
The interplay of two time scales, 
namely relaxation and renewal leads to three different phases. For small renewal rate, the  system experiences well 
separated periods of conflict (large fluctuation of $A$) followed
by long consensus state, with only minor fluctuations of  $A$, whereas for larger rates, even temporary consensus state is 
never reached and the system is constantly
pushed outwards consensus. Finally, there is a narrow transition regime in the phase space, 
in which there are numerous short periods of peace and war appearing consequently.
This three regimes are corresponding to the war scenarios discussed in Section.~\ref{sec:warscenario}. In order to make the comparison to real data 
more feasible, we defined the cumulative amount of conflict in the
system as the total sum of changes in the position of the article up to time $t$ ($t \times N$ pairs of editor-editor and editor-article interactions)
\begin{equation}
\label{eq:sumChanges}
\textstyle
S(t) = \sum_{t'=1}^t \sum_{i=1}^N | A(i) - A(i - 1) |.
\end{equation} 
In fact, the temporal evolution of $S$ shows similar patterns as 
that of $M$ obtained from Wikipedia data (Fig.\ref{fig:M-n}). For low renewal rate we have a conflict period followed 
by a peaceful one, where only minor changes happen. At large renewal rate 
we have permanent war. In between there is an alternation of conflict and peaceful regions. 
See Fig.\ref{fig:S-n} for a qualitative presentation and Ref.\cite{torok2012} for
more detailed discussion and comparison.
These results support the intuitive picture that increasing the the large rate of newcomer 
editors increase the vulnerability of the consensus.

\section{Conclusions}\label{sec:conclusion}

In this paper we surveyed recent work on WP and extended it by some new results. 
Our studies covered multilingual aspects and focused on the mechanisms and consequences of 
collaborative value production. The analysis of daily and weekly patterns of the editorial activity 
made it possible to identify the contributions from different parts of the world to such globally edited 
WPs as the English, the Spanish or the Arabic as well as to point out cultural differences in editing habits. 
The "wisdom of the crowd" seems to cope better with some tasks than pre-designed directives as the case of Simple 
WP demonstrates. Our main focus was to characterize and understand how conflicts emerge and get eventually 
resolved. While most of the WP articles are edited in a peaceful, constructive atmosphere, some of the most 
popular articles are rather controversial. In order to be able to study the conflict pages systematically, we developed 
a simple measure to identify them automatically. We have found interesting differences between peaceful and conflict 
pages in their dynamics as the edit activity of the latter is a long range correlated process in contrast to that of 
average (peaceful) pages. The language of the talk pages of conflict articles gets more reduced in complexity than that 
of regular articles and the leader-follower behavior is more intensive. 
The temporal evolution of the measure $M$ enabled to distinguish between different types of 
conflicts (single conflict with resolution, multiple conflicts, permanent war).  Finally, we showed that simple 
multi-agent modeling based on opinion dynamics can reproduce some of our findings.

%%%%%%%%%%%%%%%%%%%%%%%%%%%%%%%%%%%%%%%%%%%%%%%%%%%%%%%%%%%%%%%%%%%%%%%%%%%%%%%%%%%%
\begin{acknowledgements}
We would like to thank our collaborators: Gerardo~I\~niguez, Kimmo~Kaski, Andr\'as~Kornai, Andr\'as~Rung, 
Maxi~San~Miguel, R\'obert~Sumi, J\'anos T\"or\"ok. Discussions, advise and help with the data are gratefully 
acknowledged to Farzaneh~Kaveh, Santo~Fortunato, M\'arton~Mesty\'an, Andrzej~Nowak, Hoda~Sepehri~Rad, Attila~Zs\'{e}der, 
G\'{a}bor~Recski, Peter~Reuvern, and Katarzyna~Samson. 
\end{acknowledgements}

%%%%%%%%%%%%%%%%%%%%%%%%%%%%%%%%%%%%%%%%%%%%%%%%%%%%%%%%%%%%%%%%%%%%%%%%%%%%%%%%%%%%%


\begin{thebibliography}{100}
\providecommand{\url}[1]{{#1}}
\providecommand{\urlprefix}{URL }
\expandafter\ifx\csname urlstyle\endcsname\relax
  \providecommand{\doi}[1]{DOI~\discretionary{}{}{}#1}\else
  \providecommand{\doi}{DOI~\discretionary{}{}{}\begingroup
  \urlstyle{rm}\Url}\fi

\bibitem{aaltonen2011}
Aaltonen, A., Lanzara, G.F.: Governing complex social production in the
  internet: The emergence of a collective capability in {Wikipedia} (2011).
\newblock In Decade in Internet Time symposium

\bibitem{adler2006}
Adler, B.T., de~Alfaro, L.: A content-driven reputation system for the
  {Wikipedia}.
\newblock Tech. Rep. ucsc-crl-06-18, School of Engineering, University of
  California, Santa Cruz (2006)

\bibitem{adler2011}
Adler, B.T., de~Alfaro, L., Mola-Velasco, S., Rosso, P., West, A.: Wikipedia
  vandalism detection: Combining natural language, metadata, and reputation
  features.
\newblock In: A.~Gelbukh (ed.) Computational Linguistics and Intelligent Text
  Processing, \emph{Lecture Notes in Computer Science}, vol. 6609, pp.
  277--288. Springer Berlin / Heidelberg (2011)

\bibitem{almeida2007}
Almeida, R.B., Mozafari, B., Cho, J.: On the evolution of wikipedia.
\newblock In: Proceedings of the International Conference on Weblogs and Social
  Media, ICWSM'07 (2007)

\bibitem{apic2011}
Apic, G., Betts, M.J., Russell, R.: Content disputes in {Wikipedia} reflect
  geopolitical instabilit.
\newblock PLoS ONE \textbf{06}(6), e20,902 (2011)

\bibitem{auer2007}
Auer, S., Bizer, C., Kobilarov, G., Lehmann, J., Cyganiak, R., Ives, Z.:
  Dbpedia: A nucleus for a web of open data.
\newblock In: K.~Aberer, K.S. Choi, N.~Noy, D.~Allemang, K.I. Lee, L.~Nixon,
  J.~Golbeck, P.~Mika, D.~Maynard, R.~Mizoguchi, G.~Schreiber,
  P.~Cudré-Mauroux (eds.) The Semantic Web, \emph{Lecture Notes in Computer
  Science}, vol. 4825, pp. 722--735. Springer Berlin / Heidelberg (2007)

\bibitem{ayers2011}
Ayers, P., Priedhorsky, R.: Wikilit: collecting the wiki and wikipedia
  literature.
\newblock In: Proceedings of the 7th International Symposium on Wikis and Open
  Collaboration, WikiSym '11, pp. 229--230. ACM, New York, NY, USA (2011)

\bibitem{barabasi2005}
Barab\'asi, A.L.: The origin of bursts and heavy tails in human dynamics.
\newblock Nature \textbf{435}(7039), 207--211 (2005)

\bibitem{besten2008}
Besten, M.D., Dalle, J.: Keep it simple: A companion for {Simple} {Wikipedia}?
\newblock Industry \& Innovation \textbf{15}(2), 169--178 (2008)

\bibitem{bohannon2006}
Bohannon, J.: Tracking people's electronic footprints.
\newblock Science \textbf{314}(5801), 914--916 (2006)

\bibitem{brandes2008}
Brandes, U., Lerner, J.: Visual analysis of controversy in user-generated
  encyclopedias.
\newblock Information Visualization \textbf{7}(1), 34--48 (2008)

\bibitem{buriol2006}
Buriol, L.S., Castillo, C., Donato, D., Leonardi, S., Millozzi, S.: Temporal
  analysis of the wikigraph.
\newblock In: In Proc. of Web Intelligence, Hong Kong, pp. 45--51 (2006)

\bibitem{butler2008}
Butler, B., Joyce, E., Pike, J.: Don't look now, but we've created a
  bureaucracy: the nature and roles of policies and rules in wikipedia.
\newblock In: Proceedings of the twenty-sixth annual SIGCHI conference on Human
  factors in computing systems, CHI '08, pp. 1101--1110. ACM, New York, NY, USA
  (2008)

\bibitem{capocci2008}
Capocci, A., Rao, F., Caldarelli, G.: Taxonomy and clustering in collaborative
  systems: The case of the on-line encyclopedia {Wikipedia}.
\newblock EPL (Europhysics Letters) \textbf{81}(2), 28,006 (2008)

\bibitem{capocci2006}
Capocci, A., Servedio, V.D.P., Colaiori, F., Buriol, L.S., Donato, D.,
  Leonardi, S., Caldarelli, G.: Preferential attachment in the growth of social
  networks: The internet encyclopedia {Wikipedia}.
\newblock Phys. Rev. E \textbf{74}, 036,116 (2006)

\bibitem{zlatic2011}
Zlati\ifmmode~\acute{c}\else \'{c}\fi{}, V., \ifmmode \check{S}\else
  \v{S}\fi{}tefan\ifmmode \check{c}\else \v{c}\fi{}i\ifmmode~\acute{c}\else
  \'{c}\fi{}, H.: Model of {Wikipedia} growth based on information exchange via
  reciprocal arcs.
\newblock EPL (Europhysics Letters) \textbf{93}(5), 58,005 (2011)

\bibitem{zlatic2006}
Zlati\ifmmode~\acute{c}\else \'{c}\fi{}, V., Bo\ifmmode \check{z}\else
  \v{z}\fi{}i\ifmmode \check{c}\else \v{c}\fi{}evi\ifmmode~\acute{c}\else
  \'{c}\fi{}, M., \ifmmode \check{S}\else \v{S}\fi{}tefan\ifmmode
  \check{c}\else \v{c}\fi{}i\ifmmode~\acute{c}\else \'{c}\fi{}, H., Domazet,
  M.: Wikipedias: Collaborative web-based encyclopedias as complex networks.
\newblock Phys. Rev. E \textbf{74}, 016,115 (2006)

\bibitem{chakrabarti2006}
Chakrabarti, B.K., Chakraborti, A., Chatterjee, A. (eds.): Econophysics and
  Sociophysics: Trends and Perspectives.
\newblock Wiley-VCH Verlag GmbH \& Co, Berlin (2006)

\bibitem{cohen2010}
Cohen, J.: Computational methods for historical research on {Wikipedia's}
  archives.
\newblock e-Research \textbf{1}(2), 67--72 (2010)

\bibitem{coster2011}
Coster, W., Kauchak, D.: Simple {English} {Wikipedia}: a new text
  simplification task.
\newblock In: Proceedings of the 49th Annual Meeting of the Association for
  Computational Linguistics, HLT '11, pp. 665--669. Association for
  Computational Linguistics, Stroudsburg, PA, USA (2011)

\bibitem{mizil2012}
Danescu-Niculescu-Mizil, C., Lee, L., Pang, B., Kleinberg, J.: Echoes of power:
  Language effects and power differences in social interaction.
\newblock to appear in proceeding of WWW'12, priprint; arXiv:1112.3670  (2012)

\bibitem{deffuant2000}
Deffuant, G., Neau, D., Amblard, F., Weisbuch, G.: Mixing beliefs among
  interacting agents.
\newblock Adv. Complex Syst. \textbf{3}(4), 87--98 (2000)

\bibitem{denoyer2006}
Denoyer, L., Gallinari, P.: Acm sigir forum.
\newblock Eueopean Academy of Management Annual Conference 2010, Rome, Italy.
  \textbf{40}(1) (2006)

\bibitem{derthick2011}
Derthick, K., Tsao, P., Kriplean, T., Borning, A., Zachry, M., McDonald, D.:
  Collaborative sensemaking during admin permission granting in {Wikipedia}.
\newblock In: A.~Ozok, P.~Zaphiris (eds.) Online Communities and Social
  Computing, \emph{Lecture Notes in Computer Science}, vol. 6778, pp. 100--109.
  Springer Berlin / Heidelberg (2011)

\bibitem{ortega2009}
F., O.: Wikipedia. a quantitative analysis.
\newblock Ph.D. thesis, University Rey Juan Carlos, Madrid, Spain (2009)

\bibitem{flesch1979}
Flesch, R.: How to Write Plain English.
\newblock Harper and Row, New York (1979)

\bibitem{gabrilovich2009}
Gabrilovich, E., Markovitch, S.: Wikipedia-based semantic interpretation for
  natural language processing.
\newblock J. Artif. Int. Res. \textbf{34}, 443--498 (2009)

\bibitem{giles2005}
Giles, J.: Internet encyclopaedias go head to head.
\newblock Nature \textbf{438}, 900 (2005)

\bibitem{goh2008}
Goh, K.I., Barab\'asi, A.L.: Burstiness and memory in complex systems.
\newblock EPL \textbf{81}(4), 48,002 (2008)

\bibitem{gomez2011}
G\'{o}mez, V., Kappen, H.J., Kaltenbrunner, A.: Modeling the structure and
  evolution of discussion cascades.
\newblock In: Proceedings of the 22nd ACM conference on Hypertext and
  hypermedia, HT '11, pp. 181--190. ACM, New York, NY, USA (2011)

\bibitem{gunning1952}
Gunning, R.: The technique of clear writing.
\newblock NY: McGraw-Hill International Book Co., New York (1952)

\bibitem{gunning1969}
Gunning, R.: The fog index after twenty years.
\newblock Journal of Business Communication \textbf{6}(2), 3--13 (1969)

\bibitem{halavais2008}
Halavais, A., Lackaff, D.: An analysis of topical coverage of {Wikipedia}.
\newblock Journal of Computer-Mediated Communication \textbf{13}(2), 429--440
  (2008)

\bibitem{hardy2012}
Hardy, D., Frew, J., Goodchild, M.F.: Volunteered geographic information
  production as a spatial process.
\newblock International Journal of Geographical Information Science
  \textbf{26}(7), 1191--1212 (2012)

\bibitem{hautasaari2012}
Hautasaari, A., Ishida, T.: Analysis of discussion contributions in translated
  {Wikipedia} articles.
\newblock In: Proceedings of the 4th international conference on Intercultural
  Collaboration, ICIC '12, pp. 57--66. ACM, New York, NY, USA (2012)

\bibitem{heaps1978}
Heaps, H.S.: Information Retrieval: Computational and Theoretical Aspects.
\newblock Academic Press, Inc., Orlando, FL, USA (1978)

\bibitem{Herdan:1964}
Herdan, G.: Quantitative linguistics.
\newblock Butterworths, Washington (1964)

\bibitem{holloway2007}
Holloway, T., Bozicevic, M., Börner, K.: Analyzing and visualizing the
  semantic coverage of {Wikipedia} and its authors.
\newblock Complexity \textbf{12}(3), 30--40 (2007)

\bibitem{hu2007}
Hu, M., Lim, E.P., Sun, A., Lauw, H.W., Vuong, B.Q.: Measuring article quality
  in {Wikipedia}: models and evaluation.
\newblock In: Proceedings of the sixteenth ACM conference on Conference on
  information and knowledge management, CIKM '07, pp. 243--252. ACM, New York,
  NY, USA (2007)

\bibitem{javanmardi2009}
Javanmardi, S., Ganjisaffar, Y., Lopes, C., Baldi, P.: User contribution and
  trust in wikipedia.
\newblock In: Collaborative Computing: Networking, Applications and
  Worksharing, 2009. CollaborateCom 2009. 5th International Conference on, pp.
  1 --6 (2009)

\bibitem{javanmardi2010a}
Javanmardi, S., Lopes, C.: Statistical measure of quality in {Wikipedia}.
\newblock In: Proceedings of the First Workshop on Social Media Analytics, SOMA
  '10, pp. 132--138. ACM, New York, NY, USA (2010)

\bibitem{Javanmardi2010b}
Javanmardi, S., Lopes, C., Baldi, P.: Modeling user reputation in {Wikis}.
\newblock Statistical Analysis and Data Mining \textbf{3}(2), 126--139 (2010)

\bibitem{jones2008}
Jones, J.: Patterns of revision in online writing.
\newblock Written Communication \textbf{25}(2), 262--289 (2008)

\bibitem{jullien2012}
Jullien, N.: What we know about {Wikipedia}: A review of the literature
  analyzing the project(s) (2012).
\newblock Available at SSRN: \url{http://ssrn.com/abstract=2053597}

\bibitem{kaltenbrunner2012}
Kaltenbrunner, A., Laniado, D.: There is no deadline - time evolution of
  {Wikipedia} discussions.
\newblock In: Proceedings of the 8th International Symposium on Wikis and Open
  Collaboration, WikiSym'12. Linz (2012)

\bibitem{kampf2012}
K\"ampf, M., Tismer, S., Kantelhardt, J.W., Muchnik, L.: Fluctuations in
  {Wikipedia} access-rate and edit-event data.
\newblock Physica A: Statistical Mechanics and its Applications  (2012)

\bibitem{karkulahti2012}
Karkulahti, O., Kangasharju, J.: Surveying {Wikipedia} activity: Collaboration,
  commercialism, and culture.
\newblock In: Information Networking (ICOIN), 2012 International Conference on,
  pp. 384 --389 (2012)

\bibitem{karsai2012}
Karsai, M., Kaski, K., Barab\'asi, A.L., Kert\'esz, J.: Universal features of
  correlated bursty behaviour.
\newblock Sci. Rep. \textbf{2} (2012)

\bibitem{keegan2011}
Keegan, B., Gergle, D., Contractor, N.: Hot off the wiki: dynamics, practices,
  and structures in {Wikipedia's} coverage of the {T\=ohoku} catastrophes.
\newblock In: Proceedings of the 7th International Symposium on Wikis and Open
  Collaboration, WikiSym '11, pp. 105--113. ACM, New York, NY, USA (2011)

\bibitem{kittur2009}
Kittur, A., Chi, E.H., Suh, B.: What's in {Wikipedia}?: mapping topics and
  conflict using socially annotated category structure.
\newblock In: Proceedings of the 27th international conference on Human factors
  in computing systems, CHI '09, pp. 1509--1512. ACM, New York, NY, USA (2009)

\bibitem{kittur2008}
Kittur, A., Kraut, R.E.: Harnessing the wisdom of crowds in {Wikipedia}:
  quality through coordination.
\newblock In: Proceedings of the 2008 ACM conference on Computer supported
  cooperative work, CSCW '08, pp. 37--46. ACM, New York, NY, USA (2008)

\bibitem{kittur2007}
Kittur, A., Pendleton, B.A., Suh, B., Mytkowicz, T.: Power of the few vs.
  wisdom of the crowd: {Wikipedia} and the rise of the bourgeoisie.
\newblock In: CHI ’07: Proceedings of the SIGCHI Conference on Human Factors
  in Computing Systems (2007)

\bibitem{kornai}
Kornai, A.: Language death in the digital age.
\newblock to be published  (2012)

\bibitem{lam2011}
Lam, S.T.K., Uduwage, A., Dong, Z., Sen, S., Musicant, D.R., Terveen, L.,
  Riedl, J.: Wp:clubhouse?: an exploration of {Wikipedia's} gender imbalance.
\newblock In: Proceedings of the 7th International Symposium on Wikis and Open
  Collaboration, WikiSym '11, pp. 1--10. ACM, New York, NY, USA (2011)

\bibitem{laniado2012}
Laniado, D., Castillo, C., Kaltenbrunner, A., Fuster~Morell, M.: Emotions and
  dialogue in a peer-production community: the case of {Wikipedia}.
\newblock In: Proceedings of the 8th International Symposium on Wikis and Open
  Collaboration, WikiSym'12. Linz (2012)

\bibitem{laniado2011}
Laniado, D., Tasso, R., Volkovich, Y., Kaltenbrunner, A.: When the
  {Wikipedians} talk: Network and tree structure of {Wikipedia} discussion
  pages.
\newblock In: 5th International AAAI Conference on Weblogs and Social Media,
  ICWSM 2011, pp. 177--184 (2011)

\bibitem{lazer2009}
Lazer, D., Pentland, A., Adamic, L., Aral, S., Barab\'asi, A.L., Brewer, D.,
  Christakis, N., Contractor, N., Fowler, J., Gutmann, M., Jebara, T., King,
  G., Macy, M., Roy, D., Van~Alstyne, M.: Computational social science.
\newblock Science \textbf{323}(5915), 721--723 (2009)

\bibitem{lee2012}
Lee, J.B., Cabunducan, G., Cabarle, F.G.C., Castillo, R., Malinao, J.A.:
  Uncovering the social dynamics of online elections \textbf{18}(4), 487--505
  (2012)

\bibitem{leskovec2010}
Leskovec, J., Huttenlocher, D., Kleinberg, J.: Governance in social media: A
  case study of the {Wikipedia} promotion process.
\newblock In: Proceedings of the International Conference on Weblogs and Social
  Media, ICWSM'10 (2010)

\bibitem{leuf2001}
Leuf, B., Cunningham, W.: The {Wiki} way: quick collaboration on the {Web}.
\newblock Addison-Wesley Longman Publishing Co., Inc., Boston, MA (2001)

\bibitem{luyt2008}
Luyt, B., Aaron, T.C.H., Thian, L.H., Hong, C.K.: Improving {Wikipedia's}
  accuracy: Is edit age a solution?
\newblock J. Am. Soc. Inf. Sci. Technol. \textbf{59}(2), 318--330 (2008)

\bibitem{massa2011}
Massa, P.: Social networks of {Wikipedia}.
\newblock In: Proceedings of the 22nd ACM conference on Hypertext and
  hypermedia, HT '11, pp. 221--230. ACM, New York, NY, USA (2011)

\bibitem{masucci2011}
Masucci, A.P., Kalampokis, A., Egu\'iluz, V.M., Hern\'andez-Garc\'ia, E.:
  Wikipedia information flow analysis reveals the scale-free architecture of
  the semantic space.
\newblock PLoS ONE \textbf{6}(2), e17,333 (2011)

\bibitem{muchnik2007}
Muchnik, L., Itzhack, R., Solomon, S., Louzoun, Y.: Self-emergence of knowledge
  trees: Extraction of the {Wikipedia} hierarchies.
\newblock Phys. Rev. E \textbf{76}, 016,106 (2007)

\bibitem{napoles2010}
Napoles, C., Dredze, M.: Learning simple {Wikipedia}: a cogitation in
  ascertaining abecedarian language.
\newblock In: Proceedings of the NAACL HLT 2010 Workshop on Computational
  Linguistics and Writing, CL\&W '10, pp. 42--50. Association for Computational
  Linguistics, Stroudsburg, PA, USA (2010)

\bibitem{nielsen2011}
Nielsen, F.A.: Wikipedia research and tools: Review and comments (2011).
\newblock Available at
  \url{http://www2.imm.dtu.dk/pubdb/views/edoc_download.php/6012/pdf/imm6012.p%
df}

\bibitem{okoli2012}
Okoli, C., Mehdi, M., Mesgari, M., Nielsen, F.A., Lanam\"aki, A.: The
  people’s encyclopedia under the gaze of the sages: A systematic review of
  scholarly research on {Wikipedia} (2012).
\newblock Available at SSRN: \url{http://ssrn.com/abstract=2021326}

\bibitem{ortega2008}
Ortega, F., Gonzalez-Barahona, J., Robles, G.: On the inequality of
  contributions to {Wikipedia}.
\newblock In: Hawaii International Conference on System Sciences, Proceedings
  of the 41st Annual, p. 304 (2008)

\bibitem{park2011}
Park, T.K.: The visibility of {Wikipedia} in scholarly publications.
\newblock First Monday \textbf{16}(8) (2011)

\bibitem{pentzold2006}
Pentzold, C., Seidenglanz, S.: Foucault@wiki: first steps towards a conceptual
  framework for the analysis of wiki discourses.
\newblock In: Proceedings of the 2006 international symposium on Wikis, WikiSym
  '06, pp. 59--68. ACM, New York, NY, USA (2006)

\bibitem{ponzetto2007}
Ponzetto, S.P., Strube, M.: Knowledge derived from {Wikipedia} for computing
  semantic relatedness.
\newblock J. Artif. Int. Res. \textbf{30}, 181--212 (2007)

\bibitem{potthast2008}
Potthast, M., Stein, B., Gerling, R.: Automatic vandalism detection in
  {Wikipedia}.
\newblock In: Proceedings of the IR research, 30th European conference on
  Advances in information retrieval, ECIR'08, pp. 663--668. Springer-Verlag,
  Berlin, Heidelberg (2008)

\bibitem{ratkiewicz2010a}
Ratkiewicz, J., Flammini, A., Menczer, F.: Traffic in social media i: Paths
  through information networks.
\newblock In: Social Computing (SocialCom), 2010 IEEE Second International
  Conference on, pp. 452 --458 (2010)

\bibitem{ratkiewicz2010c}
Ratkiewicz, J., Fortunato, S., Flammini, A., Menczer, F., Vespignani, A.:
  Characterizing and modeling the dynamics of online popularity.
\newblock Phys. Rev. Lett. \textbf{105}, 158,701 (2010)

\bibitem{ratkiewicz2010b}
Ratkiewicz, J., Menczer, F., Fortunato, S., Flammini, A., Vespignani, A.:
  Traffic in social media ii: Modeling bursty popularity.
\newblock In: Social Computing (SocialCom), 2010 IEEE Second International
  Conference on, pp. 393 --400 (2010)

\bibitem{reinoso2011}
Reinoso, A.J., Gonz\a'lez-Barahona, J.M., Mu\~noz Mansilla, R.,
  Herraiz~Tabernero, I.: Temporal characterization of the requests to
  {Wikipedia}.
\newblock In: Proceedings of the 5th International Workshop on New Challenges
  in Distributed Information Filtering and Retrieval (DART 2011), vol. 771
  (2011)

\bibitem{restivo2012}
Restivo, M., van~de Rijt, A.: Experimental study of informal rewards in peer
  production.
\newblock PLoS ONE \textbf{07}(7), e34,358 (2012)

\bibitem{roth2008}
Roth, C., Taraborelli, D., Gilbert, N.: Measuring wiki viability: an empirical
  assessment of the social dynamics of a large sample of wikis.
\newblock In: Proceedings of the 4th International Symposium on Wikis, WikiSym
  '08, pp. 27:1--27:5. ACM, New York, NY, USA (2008)

\bibitem{rung}
Rung, A., Yasseri, T., Kornai, A., Kert\'esz, J.: Editorial relations in
  controversial {Wikipedia} articles.
\newblock to be published  (2012)

\bibitem{samson2010}
Samson, K., Nowak, A.: Linguistic signs of destructive and constructive
  processes in conflict.
\newblock IACM 23rd Annual Conference Paper  (2010)

\bibitem{schneider2010}
Schneider, J., Passant, A., Breslin, J.: A qualitative and quantitative
  analysis of how {Wikipedia} talk pages are used.
\newblock In: Proceedings of the WebSci10: Extending the Frontiers of Society,
  April 26-27th, 2010, Raleigh, NC: US., pp. 1--7 (2010)

\bibitem{hoda2012a}
Sepehri~Rad, H., Barbosa, D.: Identifying controversial articles in
  {Wikipedia}: A comparative study.
\newblock In: Proceedings of the 8th International Symposium on Wikis and Open
  Collaboration, WikiSym'12. Linz (2012)

\bibitem{sepehrirad2012}
Sepehri~Rad, H., Makazhanov, A., Rafiei, D., Barbosa, D.: Leveraging editor
  collaboration patterns in {Wikipedia}.
\newblock In: Proceedings of the 23rd ACM conference on Hypertext and social
  media, HT '12, pp. 13--22. ACM, New York, NY, USA (2012)

\bibitem{serrano2009}
Serrano, M.A., Flammini, A., Menczer, F.: Modeling statistical properties of
  written text.
\newblock PLoS ONE \textbf{4}(4), e5372 (2009)

\bibitem{silva2011}
Silva, F., Viana, M., Travençolo, B., da~F.~Costa, L.: Investigating
  relationships within and between category networks in {Wikipedia}.
\newblock Journal of Informetrics \textbf{5}(3), 431 -- 438 (2011)

\bibitem{smets2008}
Smets, K., Goethals, B., Verdonk, B.: Automatic vandalism detection in
  {Wikipedia}: towards a machine learning approach.
\newblock In: AAAI Workshop Wikipedia and Artificial Intelligence: an Evolving
  Synergy, WikiAI08, pp. 43--48. Association for the Advancement of Artificial
  Intelligence (2008)

\bibitem{strube2006}
Strube, M., Ponzetto, S.P.: Wikirelate! computing semantic relatedness using
  {Wikipedia}.
\newblock In: proceedings of the 21st national conference on Artificial
  intelligence, vol.~2, pp. 1419--1424. AAAI Press (2006)

\bibitem{stvilia2008}
Stvilia, B., Twidale, M.B., Smith, L.C., Gasser, L.: Information quality work
  organization in {Wikipedia}.
\newblock Journal of the American Society for Information Science and
  Technology \textbf{59}(6), 983--1001 (2008)

\bibitem{suchecki2012}
Suchecki, K., Salah, A., Gao, C., Scharnhorst, A.: Evolution of {Wikipedia's}
  category structure.
\newblock Advances in Complex Systems \textbf{15}(supp01), 1250,068 (2012)

\bibitem{suh2009}
Suh, B., Convertino, G., Chi, E.H., Pirolli, P.: The singularity is not near:
  slowing growth of wikipedia.
\newblock In: Proceedings of the 5th International Symposium on Wikis and Open
  Collaboration, WikiSym '09, pp. 8:1--8:10. ACM, New York, NY, USA (2009)

\bibitem{sumi2011a}
Sumi, R., Yasseri, T., Rung, A., Kornai, A., Kert\'esz, J.: Characterization
  and prediction of {Wikipedia} edit wars.
\newblock In: Proceedings of the ACM WebSci'11, Koblenz, Germany, pp. 1--3
  (2011)

\bibitem{sumi2011b}
Sumi, R., Yasseri, T., Rung, A., Kornai, A., Kert\'esz, J.: Edit wars in
  {Wikipedia}.
\newblock In: Privacy, Security, Risk and Trust (PASSAT), 2011 IEEE Third
  International Conference on and 2011 IEEE Third International Conference on
  Social Computing (SocialCom), pp. 724--727 (2011)
  
\bibitem{MD5}
Rivest RL (1992) The md5 message-digest algorithm.
\newblock Internet Request for Comments : RFC 1321.


\bibitem{taraborelli2010}
Taraborelli, D., Ciampaglia, G.: Beyond notability. collective deliberation on
  content inclusion in {Wikipedia}.
\newblock In: Self-Adaptive and Self-Organizing Systems Workshop (SASOW), 2010
  Fourth IEEE International Conference on, pp. 122 --125 (2010)

\bibitem{torok2012}
T\"or\"ok, J., I\~{n}iguez, G., Yasseri, T., San~Miguel, M., Kaski, K.,
  Kert\'esz, J.: Opinions, conflicts and consensus: Modeling social dynamics in
  a collaborative environment.
\newblock Phys. Rev. Lett. (in press), priprint; arXiv:1207.4914  (2012)

\bibitem{tyers2008}
Tyers, F., Pienaar, J.: Extracting bilingual word pairs from {Wikipedia}.
\newblock In: Proceedings of the SALTMIL Workshop at Language Resources and
  Evaluation Conference, LREC’08 (2008)

\bibitem{ung2010}
Ung, H.M., Dalle, J.M.: Characterizing online communities with their
  “signals”.
\newblock Eueopean Academy of Management Annual Conference 2010, Rome, Italy.
  (2010)

\bibitem{viegas2007}
Viegas, F.B., Wattenberg, M., Kriss, J., van Ham, F.: Talk before you type:
  Coordination in {Wikipedia}.
\newblock In: System Sciences, 2007. HICSS 2007. 40th Annual Hawaii
  International Conference on, p.~78 (2007)

\bibitem{volkel2006}
V\"{o}lkel, M., Kr\"{o}tzsch, M., Vrandecic, D., Haller, H., Studer, R.:
  Semantic wikipedia.
\newblock In: Proceedings of the 15th international conference on World Wide
  Web, WWW '06, pp. 585--594. ACM, New York, NY, USA (2006)

\bibitem{voss2005}
Voss, J.: Measuring {Wikipedia} (2005).
\newblock International Conference of the International Society for
  Scientometrics and Informetrics : 10th, Stockholm (Sweden), 24-28 July 2005

\bibitem{vuong2008}
Vuong, B.Q., Lim, E.P., Sun, A., Le, M.T., Lauw, H.W., Chang, K.: On ranking
  controversies in {Wikipedia}: models and evaluation.
\newblock In: Proceedings of the international conference on Web search and web
  data mining, WSDM '08, pp. 171--182. ACM, New York, NY, USA (2008)

\bibitem{wattenberg2007}
Wattenberg, M., Vi\'egas, F., Hollenbach, K.: Visualizing activity on wikipedia
  with chromograms.
\newblock In: C.~Baranauskas, P.~Palanque, J.~Abascal, S.~Barbosa (eds.)
  Human-Computer Interaction – INTERACT 2007, \emph{Lecture Notes in Computer
  Science}, vol. 4663, pp. 272--287. Springer Berlin / Heidelberg (2007)

\bibitem{west2010}
West, A.G., Kannan, S., Lee, I.: Detecting {Wikipedia} vandalism via
  spatio-temporal analysis of revision metadata.
\newblock In: Proceedings of the Third European Workshop on System Security,
  pp. 22--28

\bibitem{wiki:www}
Wikipedia: World wide web --- wikipedia{,} the free encyclopedia (2012).
\newblock
  \urlprefix\url{http://en.wikipedia.org/w/index.php?title=World_Wide_Web&oldi%
d=508583126}.
\newblock [Online; accessed 22-August-2012]

\bibitem{wilkinson2008}
Wilkinson, D.M.: Strong regularities in online peer production.
\newblock In: Proceedings of the 9th ACM conference on Electronic commerce, EC
  '08, pp. 302--309. ACM, New York, NY, USA (2008)

\bibitem{wilkinson2007}
Wilkinson, D.M., Huberman, B.A.: Assessing the value of cooperation in
  {Wikipedia}.
\newblock First Monday \textbf{12}(4) (2007)

\bibitem{wu2011}
Wu, G., Harrigan, M., Cunningham, P.: Characterizing {Wikipedia} pages using
  edit network motif profiles.
\newblock In: Proceedings of the 3rd international workshop on Search and
  mining user-generated contents, SMUC '11, pp. 45--52. ACM, New York, NY, USA
  (2011)

\bibitem{wu2010}
Wu, Q., Irani, D., Pu, C., Ramaswamy, L.: Elusive vandalism detection in
  {Wikipedia}: a text stability-based approach.
\newblock In: Proceedings of the 19th ACM international conference on
  Information and knowledge management, CIKM '10, pp. 1797--1800. ACM, New
  York, NY, USA (2010)

\bibitem{yasseri2012d}
Yasseri, T., Spoerri, A., Graham, M., Kert\'{e}sz, J.: The most controversial topics in {Wikipedia}: A multilingual
  analysis.
\newblock In preparation  (2013)

\bibitem{yasseri2012c}
Yasseri, T., Kornai, A., Kert\'{e}sz, J.: A practical approach to language
  complexity: a {Wikipedia} case study.
\newblock PLoS ONE \textbf{7}(11), e48,386 (2012)

\bibitem{mestyan2012}
Mesty\'an, M., Yasseri, T., Kert\'{e}sz, J.: Early Prediction of Movie Box Office Success based 
on {Wikipedia} Activity Big Data
\newblock Submitted, priprint; arXiv:1211.0970  (2012)

\bibitem{yasseri2012a}
Yasseri, T., Sumi, R., Kert\'{e}sz, J.: Circadian patterns of {Wikipedia}
  editorial activity: A demographic analysis.
\newblock PLoS ONE \textbf{7}(1), e30,091 (2012)

\bibitem{yasseri2012b}
Yasseri, T., Sumi, R., Rung, A., Kornai, A., Kert\'esz, J.: Dynamics of
  conflicts in {Wikipedia}.
\newblock PloS ONE \textbf{7}(6), e38,869 (2012)

\bibitem{yatskar2010}
Yatskar, M., Pang, B., Danescu-Niculescu-Mizil, C., Lee, L.: For the sake of
  simplicity: unsupervised extraction of lexical simplifications from
  {Wikipedia}.
\newblock In: Human Language Technologies 2010 Annual Conference of the North
  American Chapter of the Association for Computational Linguistics, HLT '10,
  pp. 365--368. Association for Computational Linguistics, Stroudsburg, PA, USA
  (2010)

\bibitem{zesch2008}
Zesch, T., M\"uller, C., Gurevych, I.: Extracting lexical semantic knowledge
  from {Wikipedia} and {Wiktionary}.
\newblock In: Proceedings of the Conference on Language Resources and
  Evaluation, LREC (2008)

\bibitem{zipf1935}
Zipf, G.K.: The psycho-biology of language: an introduction to dynamic
  philology.
\newblock The MIT Press, Cambridge, MA (1935)

\end{thebibliography}
\end{document}